\documentclass[fleqn,usenatbib]{mnras}

\usepackage{newtxtext,newtxmath}
\usepackage{bm}

\usepackage[T1]{fontenc}

\DeclareRobustCommand{\VAN}[3]{#2}
\let\VANthebibliography\thebibliography
\def\thebibliography{\DeclareRobustCommand{\VAN}[3]{##3}\VANthebibliography}

\usepackage{graphicx}	
\usepackage{amsmath}	
\title[Evolution mapping]{Evolution mapping: a new approach to describe matter clustering in the 
non-linear regime}

\author[A. G. S\'anchez et al.]{
Ariel G. S\'anchez,$^{1,2}$\thanks{E-mail: arielsan@mpe.mpg.de}
Andr\'es N. Ruiz$^{3,4}$
Jenny Gonzalez Jara,$^{5,6}$
and Nelson D. Padilla$^{3}$
\\
$^{1}$Max-Planck-Institut f\"ur extraterrestrische Physik, Postfach 1312, Giessenbachstr., 85748 Garching, Germany\\
$^{2}$Universit\"as-Sternwarte M\"uchen,  Fakult\"at f\"ur Physik, Ludwig- Maximilians-Universit\"at M\"unchen, Scheinerstrasse 1, 81679 M\"uchen, Germany\\
$^{3}$Instituto de Astronomía Teórica y Experimental (CONICET-UNC), Laprida 854, X5000BGR, Córdoba, Argentina\\
$^{4}$Observatorio Astronómico, Universidad Nacional de Córdoba, Laprida 854, X5000BGR, Córdoba, Argentina\\
$^{5}$Instituto de Astrofísica, Pontificia Universidad Católica de Chile, Santiago, Chile\\
$^{6}$Centro de Astro-Ingeniería, Pontificia Universidad Católica de Chile, Santiago, Chile
}

\date{Accepted XXX. Received YYY; in original form ZZZ}

\pubyear{2021}

\begin{document}
\label{firstpage}
\pagerange{\pageref{firstpage}--\pageref{lastpage}}
\maketitle

\begin{abstract}

We present a new approach to describe statistics of the non-linear matter density field 
that exploits a degeneracy in the impact of different cosmological parameters on the 
linear dimensionless matter power spectrum,  $\Delta^2_{\rm L}(k)$. We classify all 
cosmological parameters into two groups,  shape parameters, which determine the 
shape of $\Delta^2_{\rm L}(k)$,  and evolution parameters,  which only affect its amplitude 
at any given redshift. With this definition, the time evolution of $\Delta^2_{\rm L}(k)$ in 
models with identical shape parameters but different evolution parameters can be mapped 
from one to the other by relabelling the redshifts that correspond to the same clustering 
amplitude, which we characterize by the linear mass fluctuation in spheres of radius 
$12\,{\rm Mpc}$,  $\sigma_{12}(z)$.  
We use N-body simulations to show that the same evolution mapping relation gives a 
good description of the non-linear power spectrum, the halo mass function,  or the full 
density field. The deviations from the exact degeneracy are the result of the different 
structure formation histories experienced by each model to reach the same clustering 
amplitude and can be accurately described in terms of differences in the suppression 
factor $g(a) = D(a)/a$. These relations can be used to drastically reduce the number of 
parameters required to describe the cosmology dependence of the power spectrum.  
We show how this can help to speed up the inference of parameter constraints from 
cosmological observations. We also present a new design of an emulator of the non-linear 
power spectrum whose predictions can be adapted to an arbitrary choice of evolution 
parameters and redshift.
\end{abstract}




\section{Introduction}

The dramatic progress in the accuracy of cosmological observations in the last 
decades has marked the start of a data-rich era in cosmology.
These data have cemented a new cosmological paradigm, the so-called 
$\Lambda$CDM model \citep{Riess1998, Perlmutter1999, Eisenstein2005, Anderson2012, Alam2017,Planck2018, eboss2021}. 
Based on general relativity, this model is characterized by 
the presence of two components, dark matter and dark energy, whose 
nature remains elusive. 
The quest for deviations from this remarkable picture of the Universe 
will be the main objective of observational cosmology in the coming years. 

The analysis of the large-scale structure (LSS) of the Universe using data from galaxy 
surveys is one of the most effective routes to challenge the validity of the $\Lambda$CDM paradigm.
Anisotropic clustering measurements inferred from present-day galaxy redshift surveys such as 
the Baryon Oscillation Spectroscopic Survey \citep[BOSS,][]{Dawson2013} 
or the  extended Baryon Oscillation Spectroscopic Survey  \citep[eBOSS,][]{Dawson2016}
have been used to map the expansion and growth of structure histories of the Universe
over a wide range of cosmic time \citep[e.g.][]{Alam2017,eboss2021}. 
Imaging surveys with precise galaxy shape measurements offer an alternative probe of the 
LSS of the Universe. The analysis of the distortions of those shapes by the gravitational 
lensing effect caused by the foreground matter distribution has become a robust cosmological 
probe \citep{Troxel2018, Hildebrandt2018, Hikage2018, Heymans2020, DES2021}. 
Galaxy clustering and weak lensing statistics are highly complementary. Their 
joint analysis breaks the degeneracies between cosmological parameters shown by each probe, 
resulting in improved cosmological constraints
and allows for new tests that are impossible when these data sets are considered separately. 

The amount of data from galaxy surveys will increase by orders of magnitude over the 
next decade. Future surveys like the Dark Energy Spectroscopic Instrument \citep[DESI,][]{desi_survey}, 
the ESA space mission {\it Euclid} \citep{Laureijs2011}, NASA's Roman Space Telescope 
\citep{Spergel2015}, or the Legacy Survey of Space and Time (LSST) at the Rubin Observatory \citep{Ivezic2019},  
are examples of a new generation of galaxy redshift and imaging surveys that will provide more accurate 
measurements of the LSS of the Universe than ever before.
The small statistical uncertainties that are expected from these data sets 
impose demanding constraints on the accuracy of the theoretical models that will be used to 
describe them.

A basic ingredient of the analysis of LSS data sets is the modelling of 
the matter power spectrum, $P(k)$.  Galaxy clustering analyses have largely followed 
perturbation-theory based models \cite[e.g.][]{Sanchez2016a, Grieb2017, Troster2020,Damico2020,Ivanov2020}.
Although this approach has proven to be an efficient way to obtain fast an accurate 
theoretical predictions, its applicability is limited to scales in the mildly non-linear regime. 
Studies including weak lensing measurements, which have been based on small scales 
where non-linearities are strong, have mostly relied on fitting functions based on 
numerical simulations such as {\sc halofit} \citep{Smith2003, Takahashi2012, Bird2012}
or the halo model \citep{Mead2015,Mead2016,Mead2021}. 
However, the accuracy of these recipes might not be enough for future surveys. 

The analysis of the upcoming samples will require a consistent theoretical framework to describe 
galaxy clustering and weak lensing statistics that is accurate over a wide range of scales, 
including the deeply non-linear regime. 
High-resolution numerical simulations would be an ideal tool to help construct such models.
As the response of the non-linear power spectrum to changes in the cosmological parameters 
is smooth, it is possible to build accurate interpolation schemes or emulators, 
calibrated on a relatively small number of simulations 
\citep{coyote2010, Heitman2016, AemulusI, Garrison2018, Euclidemulator2018, 
Knabenhans2020, Angulo2021}. This technique can also be applied to additional 
statistics of the density field 
such as, e.g., the halo mass function \citep{Heitman2016,McClintock2019,Bocquet2020}, 
opening up the possibility to perform a battery of cosmological tests based on a consistent 
theoretical description.
However, emulators are limited by the parameter space and redshifts sampled during their 
calibration process. Currently available emulators only sample a few cosmological 
parameters, often within restrictive ranges, and are not applicable to more general parameter spaces.

In this paper,  we present a new approach to construct general models of statistics of the 
non-linear matter density field that are not restricted to a given parameter space or 
redshift range. Our method, to which we refer 
as evolution mapping, 
is based on the fact that the non-linear evolution of the density field is primarily driven  
by the linear dimensionless matter power spectrum,  
$\Delta_{\rm L}^2(k) = k^3 P_{\rm L}(k)/2\pi^2$ \citep{Hamilton1991, Peacock1994, Ma2000}.  
This means that the degeneracies between different cosmological 
parameters with respect to their impact on $\Delta_{\rm L}^2(k)$ will also be 
approximately inherited
by its non-linear counterpart.  Therefore,  the use of a parameter basis that makes 
such degeneracies more evident can be exploited to simplify the description of the 
non-linear $\Delta^2(k)$.
With this in mind,  we divide all cosmological parameters into two groups,  shape parameters,  which 
determine the shape of $\Delta_{\rm L}^2(k)$,  and evolution parameters,  which only control 
its amplitude at a given redshift $z$.
We build upon the results of \citet{Sanchez2020}, who studied the problems caused by 
the common practice of expressing theoretical predictions of cosmological measurements 
in units of $h^{-1}{\rm Mpc}$. 
\citet{Sanchez2020} showed that, when the power spectrum is expressed in Mpc units, parameters 
like the amplitude of the scalar mode, $A_{\rm s}$, and the dimensionless Hubble constant, 
$h$, follow a perfect degeneracy that defines the amplitude of $P_{\rm L}(k)$. 
As we discuss later on,  this is just a particular example of a much broader degeneracy involving all 
evolution parameters. We propose to exploit this degeneracy to describe the 
non-linear density field in terms of a reduced number of parameters. 

The structure of this paper is as follows: in Sec.~\ref{sec:parameters} we present our classification of 
cosmological parameters into shape and evolution parameters. In Sec.~\ref{sec:evmap}, we 
describe the evolution-mapping degeneracy followed by all evolution parameters and use numerical 
simulations to show that this relation, which is exact in linear theory, is also 
approximately inherited by the non-linear density field.  
In Sec.~\ref{sec:applications} we discuss a few practical applications
of evolution mapping to simplify the analysis and modelling of cosmological observables. 
Finally, Sec.~\ref{conclusions} presents our main conclusions. 
Throughout this paper, all scales are assumed to be in Mpc units and masses are expressed in 
${\rm M}_{\odot}$ without introducing any factors of $h$.

\section{Cosmological parameters and their impact on the matter power spectrum}
\label{sec:parameters}

The matter power spectrum depends on a large number of cosmological parameters describing the 
homogeneous background model and the properties of the primordial density fluctuations. 
Although different parametrisations are in principle equivalent,  the modelling 
of the non-linear power spectrum can be significantly simplified when studied in a parameter 
basis in which the degeneracies at the linear level become more explicit.

We will describe the background model in terms of the physical density parameters
\begin{equation}
\omega_{i} := \frac{8\pi G}{3\,H_{100}^2}\rho_i,
\label{eq:omegai}
\end{equation}
where $\rho_i$ represents the present-day density of each energy component $i$ (e.g. baryons, cold dark matter, 
dark energy, etc.), and $H_{100}$ is a constant given by 
\begin{equation}
H_{100} := 100\,{\rm km}\,{\rm s}^{-1}\,{\rm Mpc}^{-1},
\end{equation}
which is introduced for historical reasons and ensures that the resulting physical density parameters are 
dimensionless. 
The properties of each energy component can be described by their corresponding equation 
of state parameters
\begin{equation}
w_i := \frac{p_i}{\rho_i c^2},
\label{eq:eq_of_state}
\end{equation} 
specifying the relation between their density and pressure, which can be time dependent. 
The complete characterization of the background homogeneous model requires to specify 
its spatial curvature, $K$, which can be expressed also as a physical curvature density parameter defined as
\begin{equation}
\omega_{K} := -\frac{Kc^2}{H_{100}^2}.
\label{eq:omegak}
\end{equation}
The standard $\Lambda$CDM model corresponds to a flat Universe (i.e., $\omega_{K} =0$) 
containing contributions from photons, 
	baryons, cold dark matter, neutrinos, and dark energy, which behaves analogously to a cosmological constant 
with an equation of state parameter $w_{\rm DE}=-1$.

Besides describing the homogeneous background evolution of the Universe,
the full characterization of a cosmological model requires parameters that 
describe its inhomogeneities.
The primordial density fluctuations generated by inflation are commonly characterized in terms of the amplitude, $A_{\rm s}$,
and  the spectral index, $n_{\rm s}$, of the scalar mode at the pivot scale 
$k_{\rm p}=0.05\,{\rm Mpc}^{-1}$. 
This list is not exhaustive, as it could include additional parameters such as, e.g., the running of the
scalar spectral index $\alpha_{\rm s}$.

Multiple equivalent parameter bases could be used to specify a given cosmological model. 
For example, it is common 
to replace one of the physical density parameters (e.g. $\omega_{\rm DE}$) by the 
dimensionless Hubble parameter, $h$,  which is given by the sum of all energy 
contributions as
\begin{equation}
h^2 := \sum_i \omega_i,
\label{eq:hubble}
\end{equation}
and defines the present-day value of the Hubble parameter as $H_0 = h\,H_{100}$.
The contributions of the various energy components can also be characterized in terms 
of the density parameters 
\begin{equation}
\Omega_i := \omega_i/h^2.
\label{eq:Omegas}
\end{equation}
We will refer to the parameters $\Omega_i$ as the \emph{fractional} density parameters, 
as they represent the fraction of the total energy density of the Universe corresponding to
component $i$.
The overall amplitude of the power spectrum can be characterized by the 
RMS linear perturbation theory variance at a reference scale $R$, given by
\begin{equation}
\sigma^2(R) = \frac{1}{2\pi^2}\int{\rm d}kk^2P_{\rm L}(k)W^2(kR),
\label{eq:sigmar}
\end{equation}
where $W(kR)$ is the Fourier transform of a top-hat window of radius $R$. 
Traditionally, this reference scale has been set to $R=8\,h^{-1}\,{\rm Mpc}$. We will 
denote the corresponding value of $\sigma(R)$ as $\sigma_{8/h}$ to emphasize its dependency 
on the value of $h$. 
\citet{Sanchez2020} showed that, due to this dependency, $\sigma_{8/h}$ does not correctly capture 
the impact of $h$ on $P_{\rm L}(k)$. 
This problem can be avoided by describing the amplitude of the 
matter power spectrum in terms of a reference scale in Mpc.  As proposed by \citet{Sanchez2020}, 
a convenient choice is $12\,{\rm Mpc}$, which results in a mass variance $\sigma_{12}$ with a similar 
value to the standard $\sigma_{8/h}$ for $h\sim 0.67$. 
Alternatively, instead of $\sigma(R)$ on a given scale,  the amplitude of density fluctuations
could be characterized directly in terms of the value of the dimensionless power spectrum,  
$\Delta_{\rm L}^2(k_{\rm p})$,  at a reference wavenumber, $k_{\rm p}$,  in $\mathrm{Mpc}^{-1}$ 
\citep[see, e.g., ][]{Pedersen2021}.
We will follow \citet{Sanchez2020} and use $\sigma_{12}$ as it is closer to the most commonly
used $\sigma_{8/h}$.

To best exploit the intrinsic degeneracies between different cosmological parameters,  
we can classify them according to their impact on $\Delta_{\rm L}^2(k)$, 
into two groups, \emph{shape} and \emph{evolution} parameters. 
The first set includes the parameters that define the shape of 
the primordial power spectrum and the transfer function. Examples of these parameters are
\begin{equation}
\bm{\Theta}_{\rm s} = \left(\omega_{\gamma},\omega_{\rm b}, \omega_{\rm c}, n_{\rm s}, \cdots\right),
\label{eq:shape_param}
\end{equation}
that is, the physical densities of radiation, baryons, cold dark matter and neutrinos, 
and the scalar spectral index. 
Once the values of the parameters $\bm{\Theta}_{\rm s}$ are specified, the shape of 
the linear-theory dimensionless matter power spectrum
is completely defined.  At the linear level, all other parameters only affect the amplitude of 
$\Delta_{\rm L}^2(k)$ at any given redshift. 
These evolution parameters include 
\begin{equation}
\bm{\Theta}_{\rm e} = \left(A_{\rm s}, \omega_{K}, \omega_{\rm DE}, w_{\rm DE}(a), \cdots\right).
\label{eq:evol_param}
\end{equation}
These are the amplitude of the primordial scalar power spectrum, the curvature and dark 
energy density parameters and the dark energy equation of state parameter, including all possible 
parametrizations of its time evolution, such as the standard linear parametrization of \citet{Chevallier2001} and \citet{Linder2003}
\begin{equation}
w_{\rm DE}(a)=w_0+w_a\left(1-a\right).
\label{eq:wde}
\end{equation}
Alternatively, in early dark energy (EDE) models, which are characterized by a fractional 
dark energy density parameter that asymptotes to a value $\Omega_{\rm DE,e}$ at very high redshift, the 
time evolution of $w_{\rm DE}(a)$ can be parametrized as \citep{Wetterich2004}
\begin{equation}
w_{\rm DE}(a) = \frac{w_0}{\left(1-b\ln\left(a\right)\right)^2},
\label{eq:wa_ede}
\end{equation}
where
\begin{equation}
b = \frac{3w_0}{\ln\left(\frac{1-\Omega_{\rm DE,e}}{\Omega_{\rm DE,e}}\right)+\ln\left(\frac{1-\Omega_{\rm m}}{\Omega_{\rm m}}\right)}.
\label{eq:b_wa_ede}
\end{equation}

\begin{table}
	\centering
	\caption{Parameters of our reference flat $\Lambda$CDM model.}
	\label{tab:reference cosmology}
	\begin{tabular}{cc} 
		\hline
		Parameter & value\\
		\hline
		$\omega_{\rm b}$ &  0.02244\\
		$\omega_{\rm c} $ &  0.1206 \\
		$\omega_{\nu} $ &  0 \\
		$n_{\rm s}$ & 0.96 \\
		$\omega_K$ & 0 \\
		$\omega_{\rm DE}$ & 0.3059 \\
		$w_{\rm DE}$ & -1 \\
        $h$ & 0.67 \\
		\hline
	\end{tabular}
\end{table}

\begin{table*}
	\centering
	\caption{Test cosmologies considered in this analysis and used to define the Aletheia simulations described in Sec.~\ref{sec:evmap_nonlin}. 
	All models are characterized by identical shape parameters as in the reference cosmology defined in Table~\ref{tab:reference cosmology}
	and different evolution parameters.	 Model 0 corresponds to our reference $\Lambda$CDM universe. Models 1 to 7 are defined by changing
	one parameter of the reference case. Model 8 corresponds to an EDE cosmology. We consider the five reference values of $\sigma_{12}$ listed
	in the upper part of the table. For each cosmology, we list the redshifts at which $\sigma_{12}(z)$ matches the reference values given in the 
	upper part of the table, which define the outputs of the Aletheia simulations. 
	}
	\label{tab:all_cases}
	\begin{tabular}{lcccccc} 
		\hline
		Model & Definition&  $\sigma_{12} = 0.343$&      $\sigma_{12} = 0.499$  &    $\sigma_{12} = 0.611$  &    $\sigma_{12} = 0.703$  &   $\sigma_{12} =  0.825$\\
		\hline
		Model 0 & Reference $\Lambda$CDM as in Table~\ref{tab:reference cosmology}.& 2.000 &  1.000 & 0.570 & 0.300 & 0.00 \\
        	Model 1 & $\Lambda$CDM, $\omega_{\rm DE} = 0.2874$ ($h = 0.55$).& 1.761 & 0.859 & 0.480 & 0.248 & 0.00 \\
        	Model 2 & $\Lambda$CDM, $\omega_{\rm DE} = 0.6090$ ($h = 0.79$). & 2.231 & 1.137 & 0.659 & 0.352 & 0.00 \\
	    	Model 3 & $w$CDM, $w_{\rm DE} = -0.85$. & 2.100 & 1.044 & 0.590 & 0.307 & 0.00 \\
	    	Model 3 & $w$CDM, $w_{\rm DE} = -1.15$. & 1.923 & 0.964 & 0.553 & 0.293 & 0.00 \\
     	Model 5 & Dynamic dark energy (equation (\ref{eq:wde})), $w_a = -0.2$. & 1.973 & 0.990 & 0.566 & 0.299 & 0.00 \\
     	Model 6 & Dynamic dark energy (equation (\ref{eq:wde})), $w_a = 0.2$. & 2.031 & 1.011 & 0.574 & 0.301 & 0.00 \\
       	Model 7 & Non-flat $\Lambda$CDM, $\Omega_K = -0.05$. & 1.938 & 0.978 & 0.561 & 0.297 & 0.00 \\
       	Model 8 & EDE model, $w_0=-1.15$, $\Omega_{\rm DE,e}=10^{-5}$ . & 2.020 & 0.997 & 0.565 & 0.297 & 0.00 \\
	 	\hline
	\end{tabular}
\end{table*}

The physical density of massive neutrinos, $\omega_\nu$, deserves a special mention.  In models 
with massive neutrinos, the growth factor becomes scale-dependent even at the lineal level. Therefore, 
$\omega_\nu$ cannot be strictly classified as a shape or evolution parameter.  
We will focus on cosmologies with $\omega_\nu=0$ and leave the discussion 
on how this parameter can be included in the formalism presented here for a forthcoming analysis.

Present-day measurements of the cosmic microwave background (CMB) 
alone can accurately constrain the values of most of the 
shape parameters. However,  they only provide weak constraints on the evolution parameters
for general cosmologies.  
The inverse is true for LSS measurements. When analysed independently of 
other data sets, LSS data can only provide weak constraints on the values of the shape 
parameters. However, if the shape parameters are constrained by an independent data set, LSS 
data can provide precise measurements of the evolution parameters. 
This difference makes CMB and LSS highly complementary cosmological probes and one 
of our most powerful routes to obtain accurate and robust cosmological constraints.

\section{Evolution mapping}
\label{sec:evmap}

\subsection{Linear evolution of density perturbations}
\label{sec:evmap_lin}

As evolution parameters only change the amplitude of $\Delta_{\rm L}^2(k|z)$, 
their effect follows a perfect degeneracy.  Once a given set of shape parameters 
$\bm{\Theta}_{\rm s}$ has been specified, the power spectra of all the 
models defined by different choices of the evolution parameters $\bm{\Theta}_{\rm e}$ 
and $z$ that lead to the same clustering amplitude are identical. 
This degeneracy can be described by the value of $\sigma_{12}(z)$ as
\begin{equation}
\Delta_{\rm L}^2(k|z,\bm{\Theta}_{\rm s},\bm{\Theta}_{\rm e}) = \Delta_{\rm L}^2\left(k|\bm{\Theta}_{\rm s},\sigma_{12}\left(z,\bm{\Theta}_{\rm s},\bm{\Theta}_{\rm e}\right)\right).
\label{eq:pk_evmap}
\end{equation}   
At the linear level, the time evolution of $\Delta_{\rm L}^2(k)$ in models characterized by the 
same values of the shape parameters $\bm{\Theta}_{\rm s}$ but different choices of 
$\bm{\Theta}_{\rm e}$ can be mapped from one to the other simply by relabelling the 
redshifts that correspond to the same values of $\sigma_{12}$. 
We will therefore refer to equation~(\ref{eq:pk_evmap}) as the evolution mapping relation for 
the power spectrum. 

Note that the relation of equation~(\ref{eq:pk_evmap}) is also applicable to $P_{\rm L}(k)$ 
when it is expressed in ${\rm Mpc}$ instead of the traditional $h^{-1}{\rm Mpc}$. 
This degeneracy is also lost when the amplitude of $P_{\rm L}(k)$ is 
described in terms of $\sigma_{8/h}$, which depends on the particular value of $h$. 
The fractional density parameters $\Omega_i$ of equation~(\ref{eq:Omegas}) also 
obscure this degeneracy as they represent a mixture of shape and evolution parameters. 

\citet{Sanchez2020} discussed the perfect degeneracy between $h$ and $A_{\rm s}$ for a 
$\Lambda$CDM model, which can be characterized by a constant value of $\sigma_{12}$. 
This is a particular case of the evolution mapping relation of equation~(\ref{eq:pk_evmap}).
For a general cosmology, $h$ is a combination of shape and evolution parameters.
However, for a $\Lambda$CDM universe, fixing the values of the shape parameters 
$\omega_{\rm b}$ and $\omega_{\rm c}$ and varying $h$ corresponds to assuming different values 
of the purely evolution parameter $\omega_{\rm DE}$.

As an illustration of the relation of equation~(\ref{eq:pk_evmap}), we consider a set of nine cosmological 
models characterized by the same values of the shape parameters $\bm{\Theta}_{\rm s}$ and a wide 
range of evolution parameters $\bm{\Theta}_{\rm e}$. To define these test cosmologies, we use as a reference the 
cosmological parameters specified in Table~\ref{tab:reference cosmology}, which correspond to a flat 
$\Lambda$CDM Universe. 
Table~\ref{tab:all_cases} defines these cosmologies, labelled as models 0 to 8, 
which are specified by changing the value of one parameter of the reference case. While model 0 
corresponds to a flat $\Lambda$CDM universe close to the best-fitting model to the latest 
\textit{Planck} data (although with no contribution from massive neutrinos), the remaining cases include
different values of $h$ (corresponding to different values of $\omega_{\rm DE}$), a
time-independent dark energy equation of state $w_{\rm DE}\neq -1$, dynamical dark 
energy models with $w_a \neq 0$, a non-flat universe, and an EDE cosmology. The values of $A_{\rm s}$
of these models were defined to normalize their power spectra to give $\sigma_{12} = 0.825$ at
$z=0$. With the exception of model 0, these models were not chosen to represent viable cosmologies to describe our Universe. They represent 
extreme cases that deviate significantly from the ranges of the evolution parameters 
allowed by present-day observations. 

\begin{figure}
	\includegraphics[width=0.95\columnwidth]{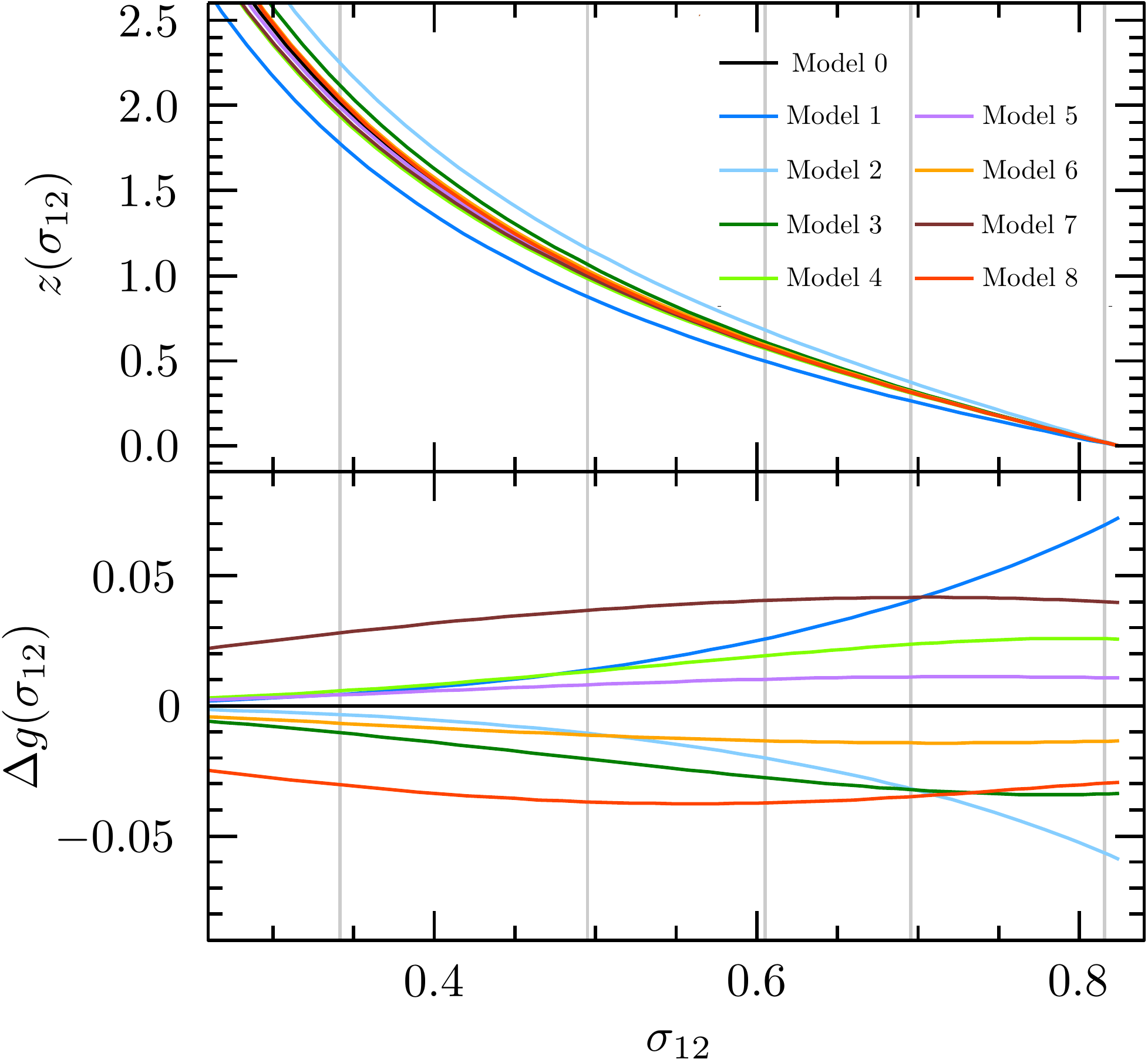}
    \caption{Upper panel: redshift at which the cosmologies defined in Table~\ref{tab:all_cases}
reach a given value of $\sigma_{12}$. Lower panel:  the differences in the suppression factors $g(a) = D(a)/a$ between 
these cosmologies and that of model 0, expressed as a function of $\sigma_{12}$.  The grey vertical lines in both panels 
correspond to the five reference values of $\sigma_{12}$ specified in the upper part of Table~\ref{tab:all_cases}. }
    \label{fig:growth_sig12}
\end{figure}

\begin{figure*}
	\includegraphics[width=0.9\textwidth]{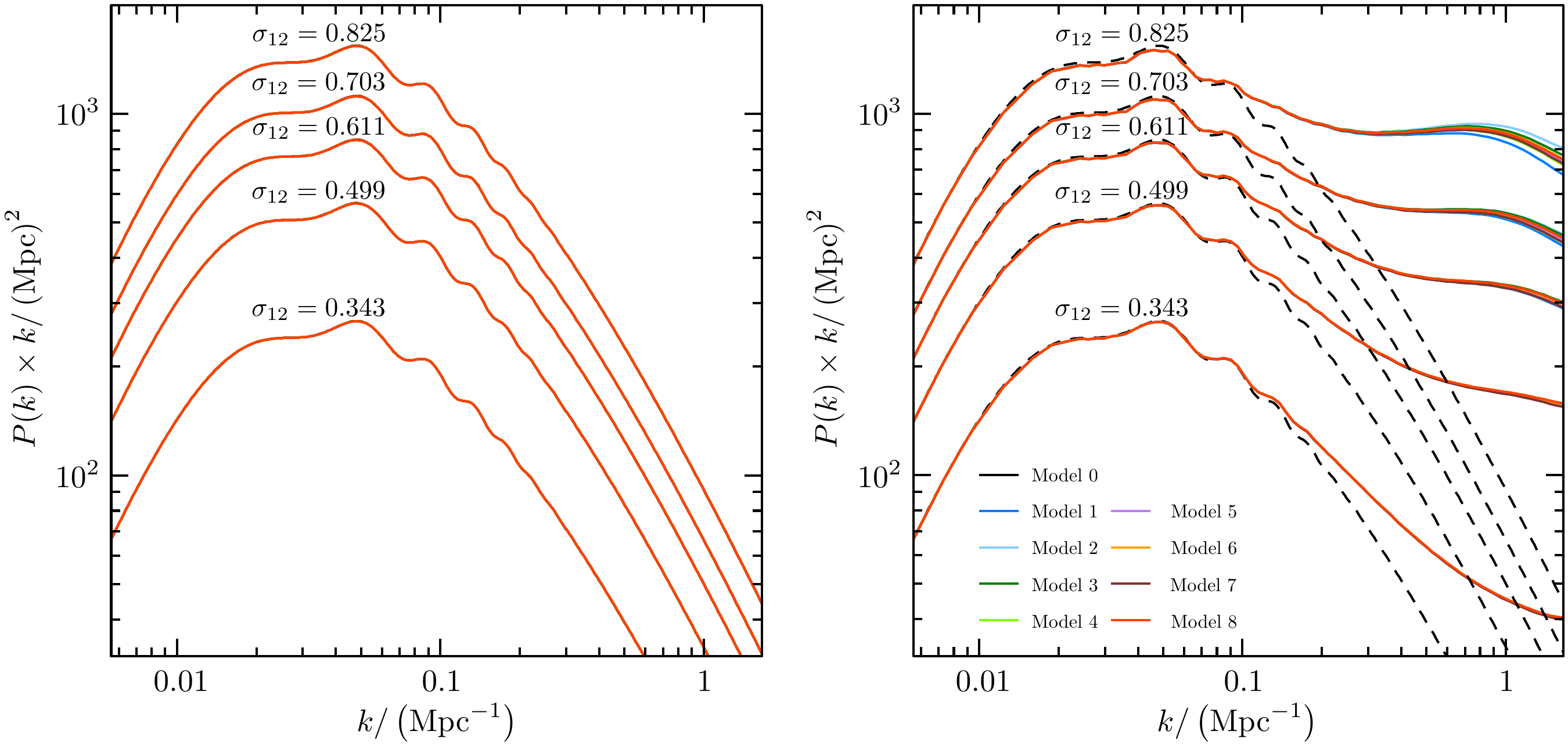}
    \caption{
Left panel: linear-theory matter power spectra computed using {\sc camb} \citep{Lewis1999} 
of the nine cosmologies defined in Table~\ref{tab:all_cases}
evaluated at the redshifts at which their values of $\sigma_{12}(z)$ match the reference values indicated 
by the labels.  Following the relation of equation~(\ref{eq:pk_evmap}), these power spectra are identical.
Right panel: matter power spectra of the same models measured from the Aletheia simulations described in 
Sec.~\ref{sec:evmap_nonlin} (solid lines) compared against their linear-theory predictions (dashed lines).  
The evolution-mapping relation of equation~(\ref{eq:pk_evmap}) continues to give 
a good description of the results.}
    \label{fig:pk_only}
\end{figure*}

The upper panel of Fig.~\ref{fig:growth_sig12} shows the redshift at which these test cosmologies reach a
given value of $\sigma_{12}$. Equation~(\ref{eq:pk_evmap}) implies that the linear power spectra 
of these models will be identical when evaluated at the redshifts that correspond to the same value of  
$\sigma_{12}$. As an illustration of this relation, we used the five values of $\sigma_{12}$ 
specified in the upper part of Table~\ref{tab:all_cases}, which are indicated by vertical grey lines in 
Fig.~\ref{fig:growth_sig12}.
The redshifts at which the value of $\sigma_{12}(z)$ for each model matches these reference values 
are also listed 
in Table~\ref{tab:all_cases}. As can be seen in the left panel of Fig.~\ref{fig:pk_only}, 
when they are evaluated at these redshifts, 
the linear-theory power spectra of these models are indistinguishable. 

Although we have focused on the dark matter power spectrum in real space, a relation
equivalent to equation~(\ref{eq:pk_evmap}) will also be valid for biased tracers in redshift 
space. In this case, the power spectrum will depend on the linear bias parameter, $b(z)$,  
and the logarithmic growth rate $f(z)=d \ln D(z)/d \ln a$, with $D(z)$ the linear growth
factor and $a$ the scale factor of the Universe \citep{Kaiser1987}.  
At the linear level, all models characterized by identical shape parameters and 
the same values of the parameter combinations $b\sigma_{12}(z)$ and $f\sigma_{12}(z)$ 
will be identical. For this reason, as discussed by \citet{Sanchez2020}, the combination 
$f\sigma_{12}(z)$ gives 
a more correct description of the cosmological information content of the pattern of 
redshift-space distortions than the commonly used $f\sigma_{8/h}(z)$.  
We leave the analysis of the evolution mapping relation in redshift space and its interplay 
with the geometric Alcock-Paczynski distortions for future work.

\subsection{The non-linear matter power spectrum}
\label{sec:matterpk}
\label{sec:evmap_nonlin}

 It is well known that the non-linear evolution of the matter power spectrum is mainly determined 
by its linear-theory counterpart $\Delta^2_{\rm L}(k|z)$ 
\citep[e.g.,][]{Hamilton1991, Peacock1994,Peacock1996,Jain1995,Ma2000}. This fact is 
at the core of some of the most commonly used tools to account for non-linearities 
in $P(k)$  \citep[such as, e.g.,  {\sc halofit},][]{Smith2003},  
and can be understood in the context of standard perturbation theory (SPT) and other related
approaches.  Assuming that the perturbation theory kernels are independent of cosmology, 
which has been shown to be a good approximation even for non-standard cosmologies
\citep{Takahashi2008,Taruya2016,Garny2021},  SPT implies that the non-linear $P(k)$ is a function 
 of the linear power that is independent of the cosmological parameters \citep{Scoccimarro1998}.  
As an ilustration of this behaviour,  we can use renormalized perturbation theory  
\citep[RPT;][]{Crocce2006},  in which the non-linear matter power spectrum,  $P(k|z)$,  
can be written as 
\begin{equation}
P(k|z)=P_{\rm L}(k|z)\, G(k|z)^2+P_{\rm MC}(k|z),
\label{eq:pk_rpt}
\end{equation}
where the propagator $G^2(k|z)$ is obtained by resumming all the terms in the 
standard perturbation theory expansion 
that are proportional to the linear power spectrum $P_{\rm L}(k|z)$, and 
$P_{\rm MC}(k|z)$ contains 
all mode-coupling contributions. The first term represents the contribution to the final $P(k|z)$
coming from the linearly-evolved power at the same scale $k$, while the second one
describes the contribution from all other scales in the linear power. 
The propagator is given by a nearly Gaussian damping, whose characteristic scale is 
defined by an integral over the linear power spectrum. 
The mode-coupling term can be expressed as a sum of a series of loop contributions, which at 
$N$ loops involve convolutions over $N$ linear power spectra. 
Hence, for models with identical $P_{\rm L}(k|z)$, RPT will also lead to the same predictions 
for the propagator and mode coupling terms, leading to indistinguishable non-linear power
spectra. This also applies to other commonly used 
recipes to describe the non-linear power spectrum that depend exclusively 
on $P_{\rm L}(k|z)$ \citep[e.g.][]{Taruya2012,Nishimichi2017}.

Equation~(\ref{eq:pk_evmap}) provides us with a practical recipe to map the evolution of 
models characterized by identical shape parameters but different evolution parameters 
that is exact at the level of linear perturbations.  
In the context of perturbation theory,  the same mapping would be applicable 
to the power spectrum in the non-linear regime.  
However, the fundamental assumption of single-stream flow of common perturbation 
theory approaches eventually breaks down due to shell crossings on small scales. 
Models with the same $P_{\rm L}(k|z)$ but different structure growth histories 
show different non-linear power spectra \citep{Mead2017}. 
Therefore, there will be deviations from equation~(\ref{eq:pk_evmap}) in the deeply 
non-linear regime.

To test this in detail, we ran numerical simulations corresponding to the models listed in 
Table~\ref{tab:all_cases}. For each model, we used {\sc gadget}-4 \citep{Springel2021}
to generate two simulations following the fixed-paired approach to suppress 
cosmic variance of \citet{Angulo2016}. Each simulation 
followed the evolution of $1500^3$ dark matter particles on a box of side
$L_{\rm box} = 1492.5\,{\rm Mpc}$. 
The simulations were started at redshift $z=99$ from initial conditions 
generated with {\sc 2LPTic} \citep{crocce_2lptic}, using the same random 
phases for all models.
Both {\sc 2LPTic} and {\sc gadget}-4 were modified to include different dark energy models.  
The Plummer-equivalent softening length was set to $22\,{\rm kpc}$, corresponding to 
2 per-cent of the mean inter-particle separation.
Each simulation has 5 snapshots chosen to match the redshifts at which each model reaches the 
reference values of $\sigma_{12}$ listed in Table~\ref{tab:all_cases}. 
We refer to this set as the \textit{Aletheia} simulations.

\begin{figure}
	\includegraphics[width=\columnwidth]{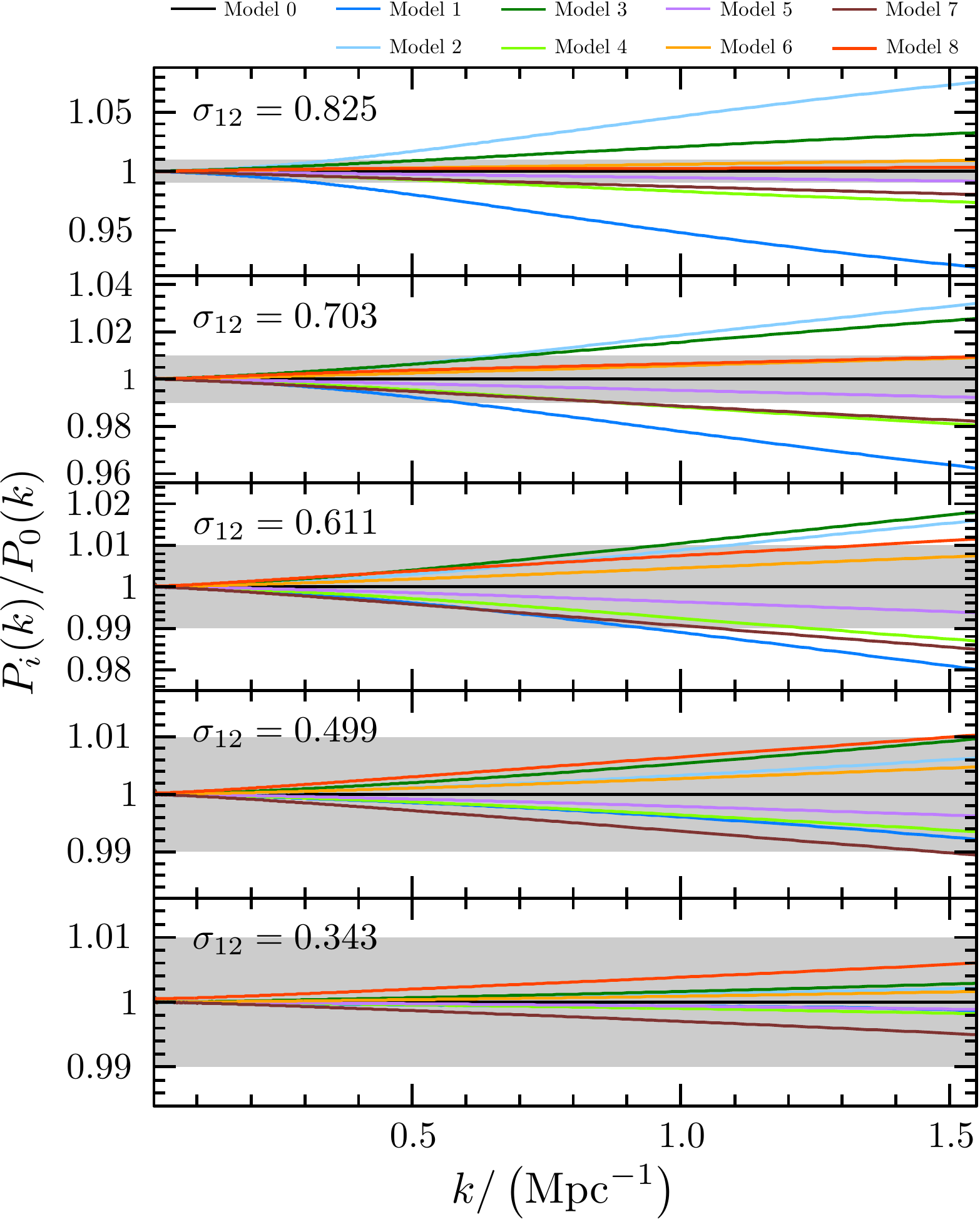}
    \caption{Ratios of the matter power spectra recovered from the different Aletheia simulations 
    and the one corresponding to model 0 for each of our reference values of $\sigma_{12}$ in 
    different panels.}
    \label{fig:pk_ratios}
\end{figure}

\begin{figure}
	\includegraphics[width=\columnwidth]{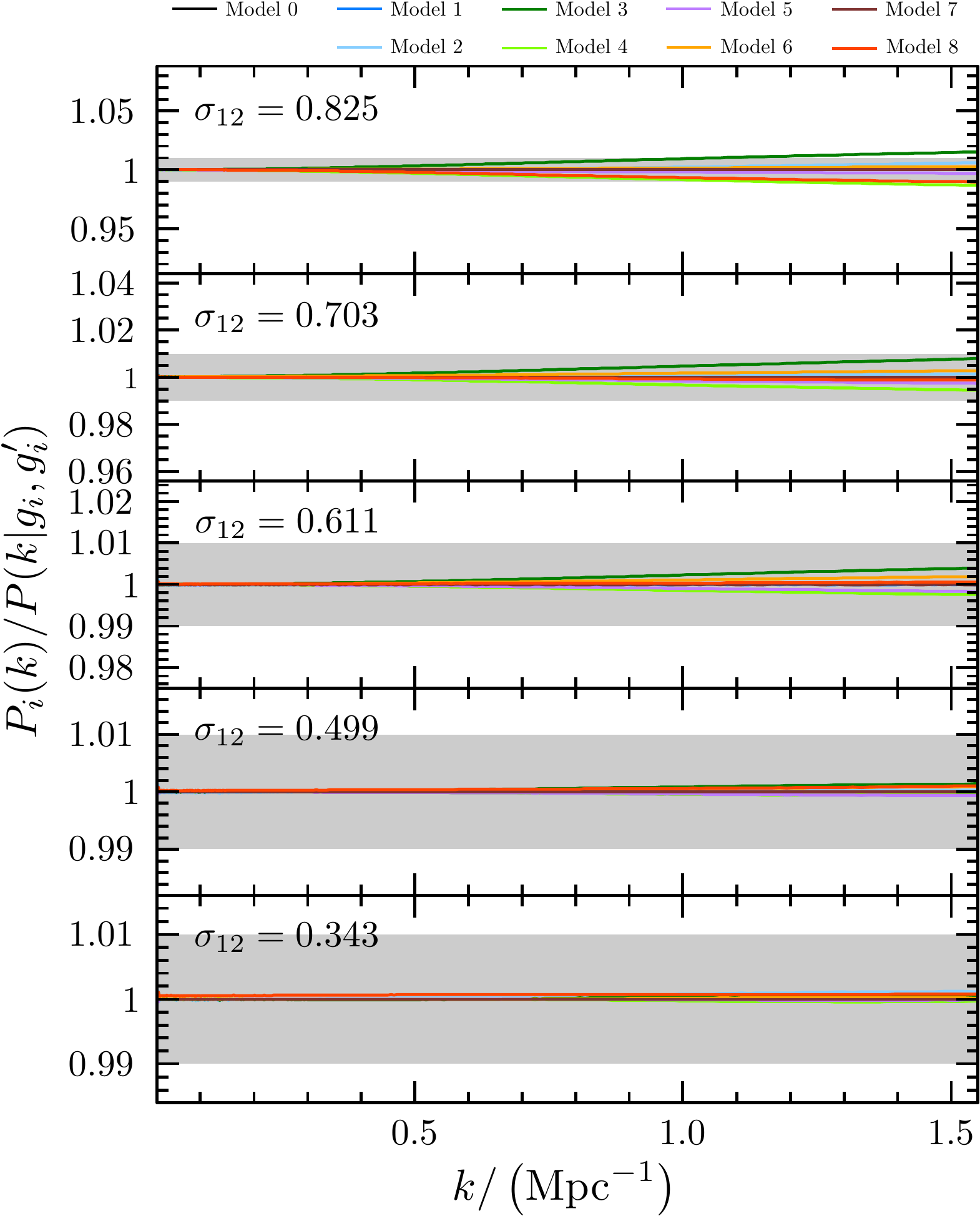}
    \caption{Ratios of the matter power spectra recovered from model $i$ of the Aletheia simulations 
    and the ones predicted using equation~(\ref{eq:pk_taylor}) based on the results of model 0 and the
    differences in the growth of structure histories of these models, characterized by the values of 
    $\Delta g_i$ and $\Delta g'_i$, for each reference value of $\sigma_{12}$.}
    \label{fig:pk_ratios_der}
\end{figure}

We computed the matter power spectra of each snapshot of all simulations 
using the available option in {\sc gadget}-4 and averaged the 
measurements from each pair corresponding to the same model.
We focus on wavenumbers $k < 1.5\,{\rm Mpc}^{-1}$
to avoid scales where baryonic effects, which we are ignoring in our analysis, 
would have to be taken into account.
The solid lines in the right panel of Fig.~\ref{fig:pk_only} show the resulting
power spectra, which exhibit clear deviations from their linear-theory
predictions, shown by the black dashed lines. 
The evolution mapping relation of equation~(\ref{eq:pk_evmap})
continues to give a very good description of the results. 
Despite the wide range of evolution parameters covered by these models, their 
power spectra are in good agreement when they are evaluated at the
redshifts that correspond to the same values of $\sigma_{12}$. 

As expected, unlike the naive expectation based on perturbation theory, the relation of 
equation~(\ref{eq:pk_evmap}) is not exact in the deeply non-linear regime. 
The differences between these models can be seen more clearly in Fig.~\ref{fig:pk_ratios}, 
which shows the ratios of the power spectra of all models with respect to that of model 0.
The differences increase with $k$, and are larger for higher values of $\sigma_{12}$. 
As a reference, the grey shaded areas shown in all panels correspond to a 1 per-cent difference. 
These differences with the power spectrum of model 0 remain at the sub-percent 
level for $\sigma_{12} \leq 0.499$. The maximum differences are seen at 
$\sigma_{12} = 0.825$, corresponding to $z=0$ in all models, and can reach an
8 per-cent level at $k = 1.5\, {\rm Mpc}^{-1}$. 
Note also that the models that deviate the most from model 0 
are different at each value of $\sigma_{12}$. For example, while models 1 and 2 
exhibit the most significant deviations for $\sigma_{12} \geq 0.703$, models 7 and 8 
show the largest differences for $\sigma_{12} < 0.611$.
Despite these differences, the power spectra ratios show no leftover from the signature of 
baryon acoustic oscillations, indicating that the damping of this signal with respect to 
the linear-theory prediction is the same in all cosmologies.
Appendix~\ref{sec:appendix} presents a comparison of these results 
with the predictions of available recipes to model the non-linear $P(k|z)$.

Previous analyses \citep{McDonald2006, Ma2007,Mead2017} have 
studied the differences in the non-linear power
spectrum between cosmological models characterized by the same $P_{\rm L}(k|z = 0)$ 
but different growth of structure histories.  Focusing on 
$\Lambda$CDM cosmologies and models with $w_{\rm DE}\neq -1$ and identical 
values of $h$, 
these studies have found that their non-linear power spectra at $z=0$ are also approximately 
the same in the mildly non-linear regime but differ at smaller scales.  Our results extend those
findings to general cosmologies characterized by the same shape parameters and 
any choice of evolution parameters, as long as they are compared at the redshifts 
at which their values of $\sigma_{12}(z)$ are identical. This is also valid for 
cosmologies with different values of $h$ as long as the power spectra are expressed in 
${\rm Mpc}$ units.

\begin{figure*}
	\includegraphics[width=0.95\textwidth]{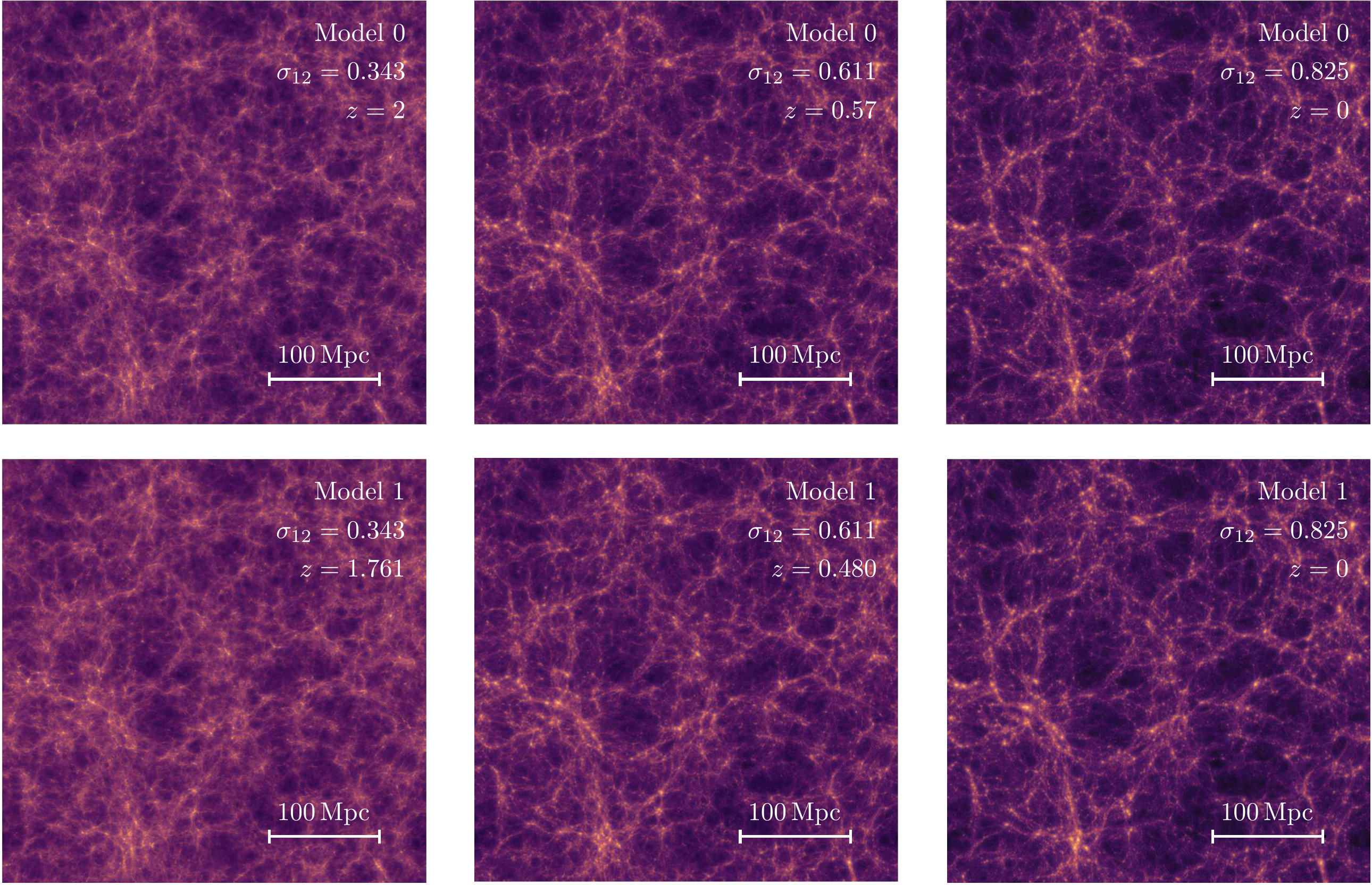}
    \caption{Section of the density field projected over a slice 75 Mpc wide of one of the Aletheia 
    simulations of model 0 (upper panels) and the equivalent one for model 1 (lower panels) 
    at the redshifts corresponding to three of our reference values of $\sigma_{12}$. 
    The full pattern of structures such as voids, filaments, and haloes is reproduced with striking 
    agreement in the two realizations. Note that this agreement would be lost if the particle positions
    were expressed in the commonly used $h^{-1}{\rm Mpc}$ units as the value of $h$ is different in 
    these two models. }
    \label{fig:slices}
\end{figure*}

Already in the recipe of \citet{Peacock1994} the deviations between the non-linear 
power spectra of models with identical $\Delta^2_{\rm L}(k)$ caused by their different 
structure formation histories are described in terms of the suppression factor $g(a) = D(a)/a$.
The lower panel of Fig.~\ref{fig:growth_sig12} shows the differences in the suppression 
factors $g(a)$ between the different Aletheia cosmologies  
and that of model 0, expressed as a function of the corresponding value of $\sigma_{12}$. 
These differences show a similar structure to the deviations between the non-linear power spectra 
shown in Fig.~\ref{fig:pk_ratios}. For each of our reference values of $\sigma_{12}$, the models with the largest differences 
$\Delta g(\sigma_{12})$ are the ones for which $P(k)$ exhibits the largest deviations from that of model 0. 
With this in mind, we tested a simple ansatz to describe the residuals with respect to the 
evolution mapping relation of equation~(\ref{eq:pk_evmap}) in terms of $g(\sigma_{12})$ 
and $g'(\sigma_{12}) = dg(\sigma_{12})/d \sigma_{12}$ as 
\begin{equation}
\begin{split}
P\left(k|g,g'\right) =&\,\, P\left(k|g_0, g'_0\right) + 
\frac{\partial P }{\partial g}\left(k|g_0, g'_0\right)\left(g - g_0\right) \\ 
&+ 
\frac{\partial P }{\partial g'}\left(k|g_0, g'_0\right)\left(g' - g'_0\right),
\label{eq:pk_taylor}    
\end{split}
\end{equation}
where for simplicity we omitted the dependency on $\bm{\Theta}_{\rm s}$ and $\sigma_{12}$, 
which are kept fixed in all terms. To test this ansatz on the Aletheia simulations, we used 
 model 0 as a reference and inferred the derivatives with respect to 
$g$ and $g'$ using the power spectra of models 1 and 7. 
We then used these results in equation~(\ref{eq:pk_taylor}) to compute predictions for the 
non-linear power spectra of all other models. Fig.~(\ref{fig:pk_ratios_der}) shows the ratios of 
the matter power spectra of all Aletheia 
simulations and their predictions based on equation~(\ref{eq:pk_taylor}). By construction, 
this relation gives a 
perfect match to the power spectra of models 0, 1, and 7 but it also gives a good description 
of the results of all other models, with deviations that remain smaller than 1 per-cent in 
almost all cases. 
Despite its simplicity, equation~(\ref{eq:pk_taylor}) captures the main impact of the 
different structure formation histories
of the  Aletheia cosmologies on their respective non-linear power spectra. 
As we will see in the next section, the same approach can be applied to describe 
other statistics of the non-linear matter density field. 

It is interesting to consider our results in the context of the cosmology 
rescaling of \citet{Angulo2010}.  For cosmologies with identical linear power spectrum 
shape,  this rescaling reduces to a relabelling of the redshifts to match the 
global amplitude of density fluctuations as in the relation of equation~(\ref{eq:pk_evmap}).
The additional step of rescaling of the mass -- concentration relation of \citet{Contreras2020}
is analogous to the correction for the impact of the different structure formation histories 
in terms of $g(\sigma_{12})$ of equation~(\ref{eq:pk_taylor}). 
This picture is consistent with the results of \citet{Diemer2019} who found that,  although 
the overall amplitude and shape of $P_{\rm L}(k|z)$ (characterized by the peak height 
parameter, $\nu \propto \sigma^{-1}$,  and the effective power spectrum slope, $n_{\rm eff}$)
can account for most of the cosmology dependence of the concentration -- mass relation,  a 
more accurate description requires also information of the growth-rate of cosmic structure in
the different models.

\subsection{Evolution mapping beyond two-point statistics}
\label{sec:nm}

\begin{figure}
	\includegraphics[width=\columnwidth]{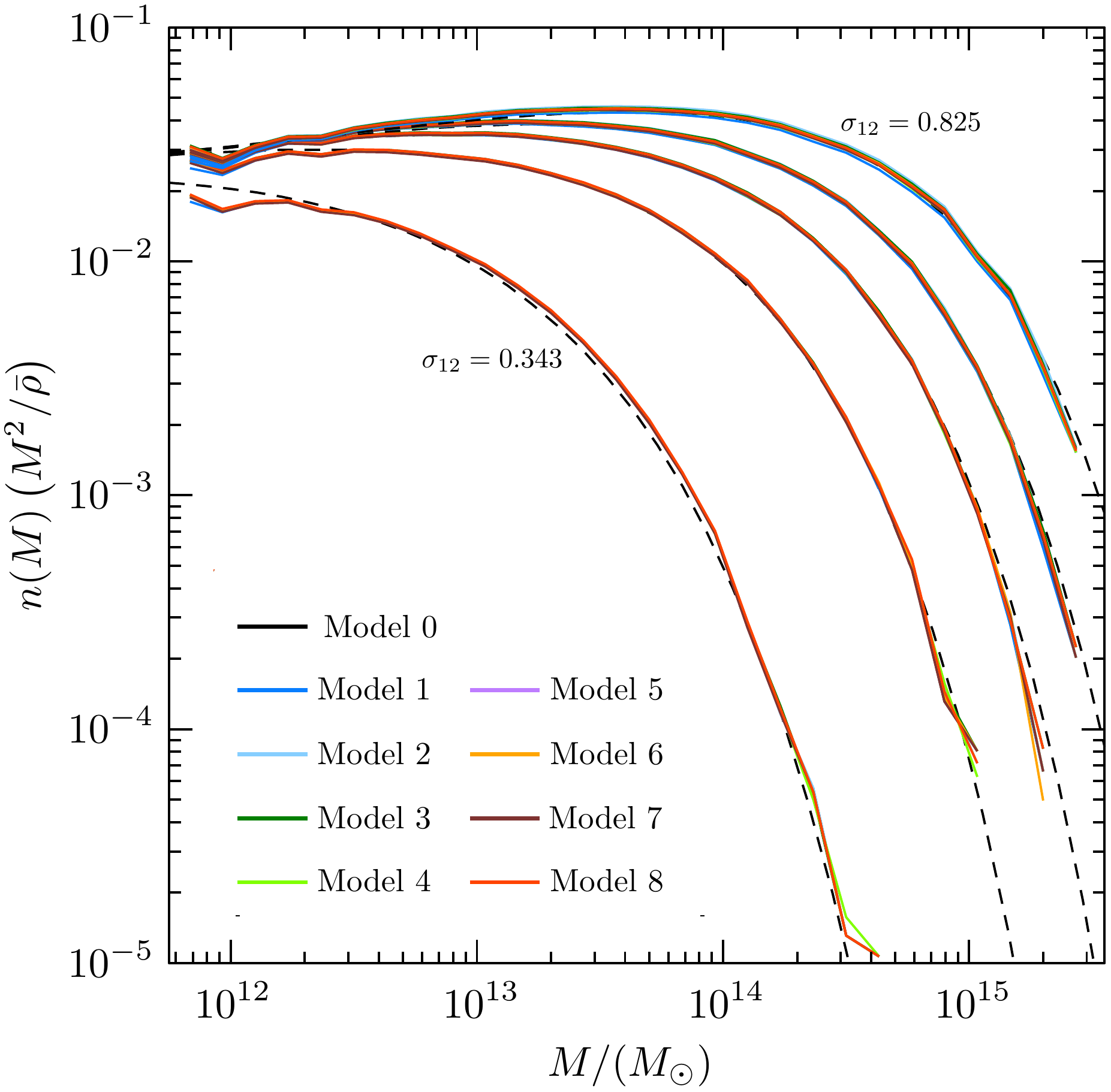}
    \caption{
Halo mass functions measured from the Aletheia simulations 
(solid lines) for the snapshots that correspond to our five reference values of $\sigma_{12}(z)$. 
The prediction of \citet{Tinker2008} for model 0 (dashed lines) gives a good description of these 
measurements.	The agreement between these measurements is lost when halo masses are expressed in 
$h^{-1}M_\odot$ units.}
    \label{fig:nm_all}
\end{figure}

The upper panels of Fig.~\ref{fig:slices} show slices over the box of one of the Aletheia 
simulations of model 0, at $z=2$, 0.57, and 0, 
corresponding to three of our reference values of $\sigma_{12}$. 
The lower panels show slices of the same region of the realization 
of model 1 with matching initial condition phases, at  $z=1.761$, 0.48, 
and 0, which correspond to the same values of  $\sigma_{12}$. 
These figures illustrate that, when the growth of structure is traced using the 
global clustering amplitude as a reference,  the full density field with its variety of 
structures is reproduced with striking agreement.

The initial conditions of these simulations differ only in the amplitude of their power spectra.
\citet{Nusser1998} showed that the equations of motion of a system of collisionless 
particles can be expressed in a form that is almost independent of the 
cosmological parameters when the evolution is described in terms of the variable
$\tau = \ln\,D$. 
Neglecting this weak dependence, the simulations of the two models will follow 
exactly the same evolution but with a given amplitude of density fluctuations taking 
place at different values of $\tau$ in each case. 
Using instead the variable $\tau' = \ln\,\sigma_{12}$, which corresponds to a constant 
shift in $\tau$ for each cosmology, synchronizes both solutions. 
The weak cosmology dependence of the equations of motion leads to small differences 
in the two solutions, which become evident in the high-density regions that dominate
the power spectrum at high $k$ values.

The similarity illustrated in Fig.~\ref{fig:slices} suggests that evolution mapping does 
not only apply to the matter power spectrum. A relation similar to equation~(\ref{eq:pk_evmap})
must be valid for higher-order $N$ point statistics, geometrical and topological
descriptors such as Minkowski functionals, or the full probability distribution
function of the density field. We leave a detailed study of such relations for future work. 
We focus here on another important source of information of the non-linear matter 
density field, the halo mass function, $n(M)$.

Most  theoretical models of the halo mass function are expressed in terms of the 
halo multiplicity function given by 
\begin{equation}
f(\sigma) = \frac{M}{\bar{\rho}(0)}\frac{d n(M)}{d \ln\left(\sigma^{-1}\right)},
\label{eq:fsigma}
\end{equation}
where $\sigma^2(M)$ represents the variance of the linearly-evolved density field 
given in equation~(\ref{eq:sigmar}) evaluated at a scale 
\begin{equation}
R(M) = \left(\frac{3\,M}{4\pi\bar{\rho}}\right)^{1/3}.
\end{equation}
Theoretical recipes based on the spherical or ellipsoidal collapse models 
predict that $f(\sigma)$ is independent of cosmology and redshift \citep{Press1974,Sheth1999}. 
Using N-body simulations, \citet{Jenkins2001} found that the function $f(\sigma)$ corresponding 
to different halo definitions is close to universal and provided a fitting function
accurate at the 10-20 per-cent level.
The same strategy has been followed by several authors \citep{Reed2003, Warren2006,Reed2007,
Tinker2008,Crocce2010, Watson2013,Bocquet2016,Seppi2020}, leading to prescriptions with 
improved accuracy.
In the context of these recipes,  cosmologies with identical linear power spectra, and hence
equal variance $\sigma^2(M)$, would also have indistinguishable mass functions, implying that a 
relation analogous to equation~(\ref{eq:pk_evmap}) should also be valid, 
at least approximately,  for $n(m)$. 

To study the applicability of an evolution mapping relation to the mass function, 
we computed $n(M)$ for the Aletheia simulations.
Different definitions of halo mass, $M$, have been used in the literature primarily
based on the friends-of-friends (FOF) percolation algorithm \citep{Davis1985} or on
the spherical overdensity (SO) halo finder \citep{Lacey1994}, 
with the latter having a more direct link to theoretical predictions. 
SO masses are often defined as a given overdensity with respect to the critical density
\begin{equation}
\rho_{\rm c}(z) = \frac{3\,H(z)^2}{8\pi G}.
\end{equation} 
However, this definition depends explicitly on the Hubble parameter which, 
as discussed in Sec.~\ref{sec:parameters}, represents a mixture of shape and 
evolution parameters and would therefore spoil the possible use of an evolution
mapping relation for $n(M)$.
Instead, we used SO masses defined by the radius enclosing an average density 
that is a factor $\Delta=200$ of the mean density of the 
Universe, $\bar{\rho}(z=0)$.

\begin{figure}
	\includegraphics[width=\columnwidth]{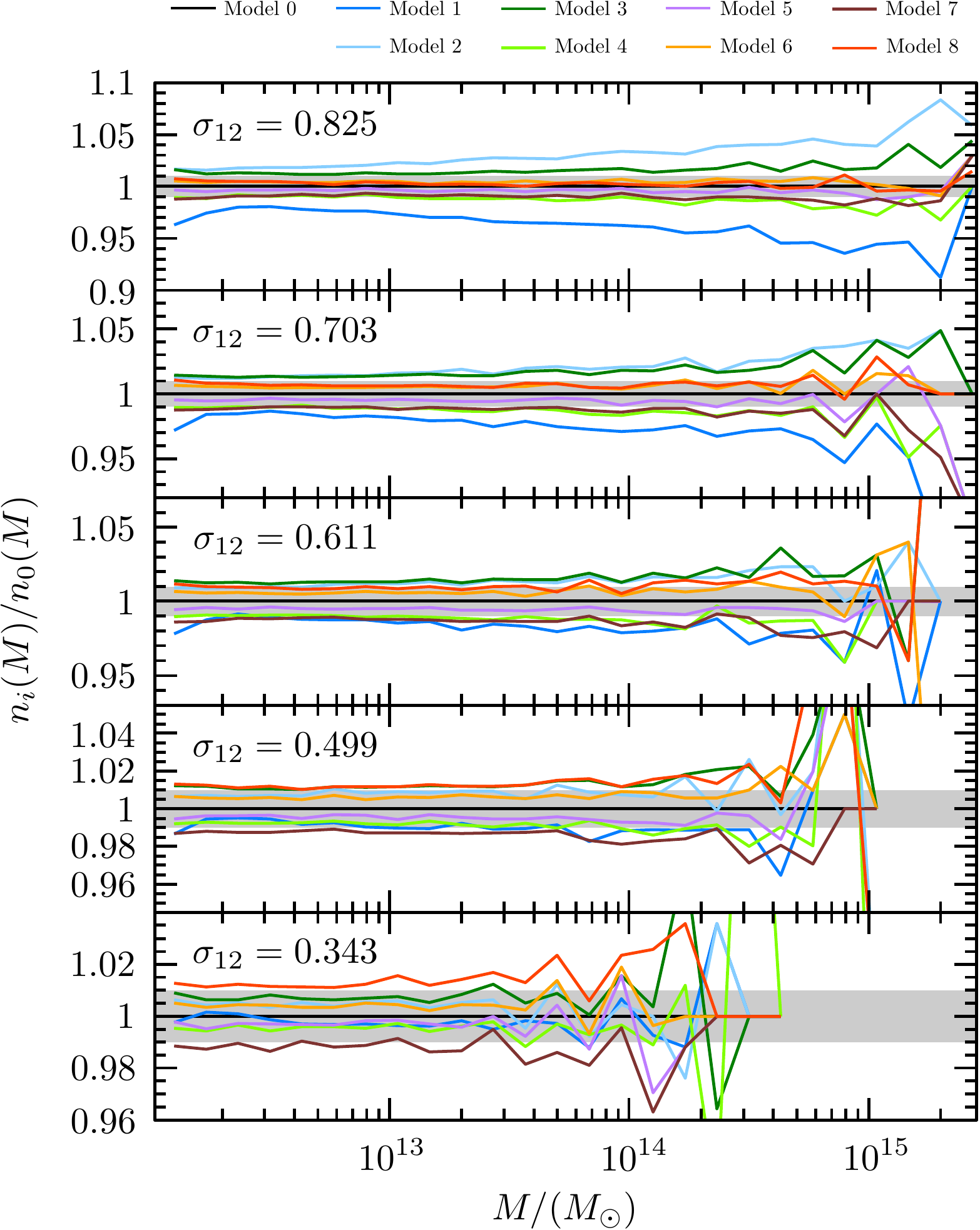}
    \caption{Ratios of the halo mass functions recovered from the different Aletheia simulations 
    and the one corresponding to model 0 for our five reference values of $\sigma_{12}$.}
    \label{fig:nm_ratios}
\end{figure}

We identified dark matter haloes and their properties using {\sc rockstar} \citep{Rockstar} 
and computed the mass functions of all Aletheia simulations averaging the 
measurements from the pairs corresponding to the same cosmological model.
Fig.~\ref{fig:nm_all} shows the mass functions of the models for our reference 
values of $\sigma_{12}$ plotted as $n(M)(M^2/\bar{\rho}^2(0))$, 
which are remarkably similar. Note that this agreement would have been hidden for models 1 and 2 if 
the measurements were expressed in the commonly used units of $h^{-1}{\rm M}_\odot$ 
and $h^{-1}{\rm Mpc}$ as these cosmologies are characterized by different values of $h$.
The black dashed lines show the predictions for the mass function of model 0 
at each redshift based on the recipe of \citet{Tinker2008}, which is in good agreement with our
measurements.
The differences between the various models can be seen more clearly in Fig.~\ref{fig:nm_ratios}, 
which shows the ratios of the mass functions of all Aletheia simulations, with respect to that 
of model 0. 
These ratios show similar trends as those of $P(k)$, with sub-percent level differences
at low values of $\sigma_{12}$ that increase to a few percent for $\sigma_{12}=0.825$. 

\begin{figure}
	\includegraphics[width=\columnwidth]{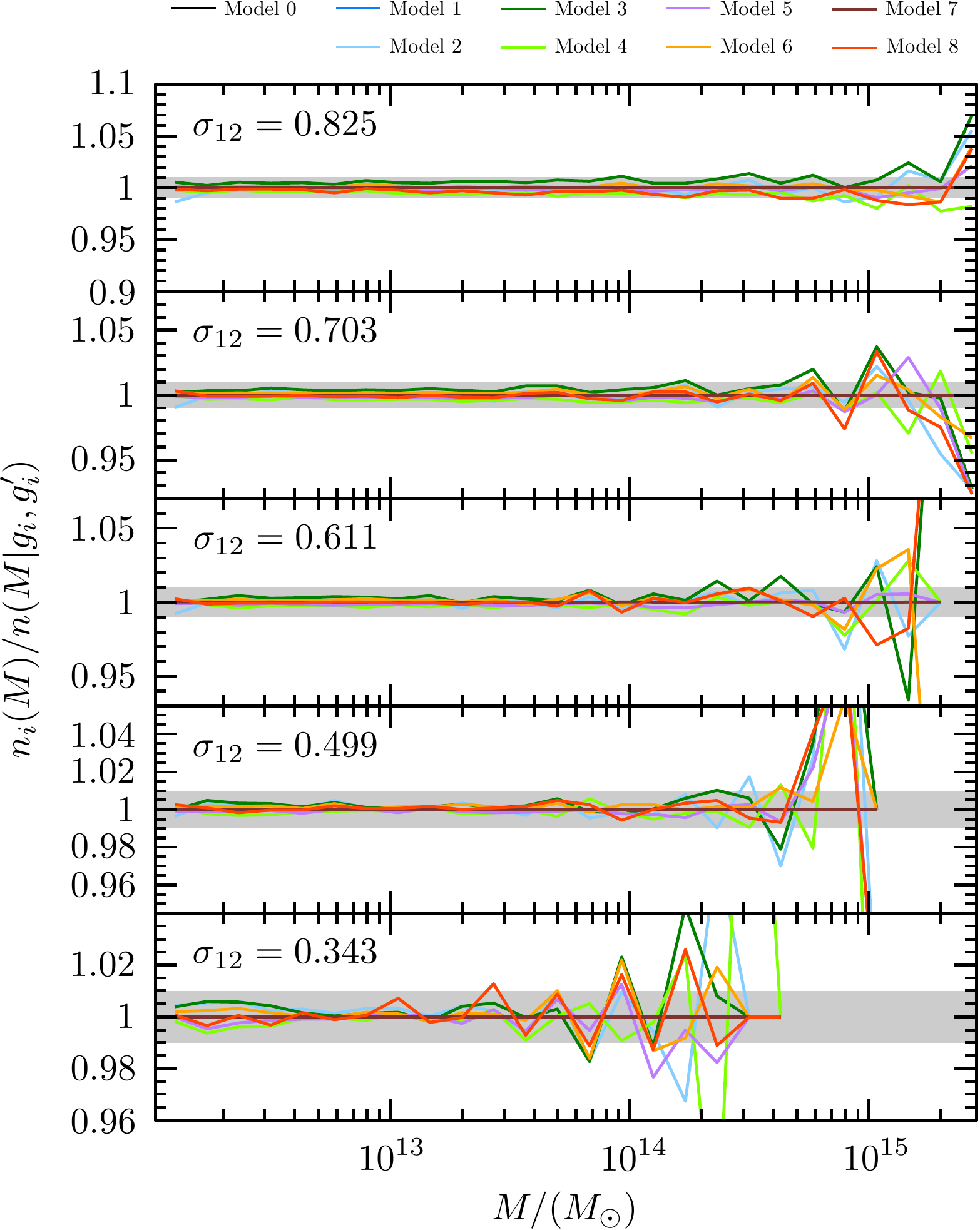}
    \caption{	
    Ratios of the halo mass functions recovered from the different models $i$ of the Aletheia simulations 
    and the ones predicted using equation~(\ref{eq:nm_taylor}) based on the results of model 0 and the
    differences in the growth suppression factors, $\Delta g_i$, and its derivative with respect to 
    $\sigma_{12}$, $\Delta g'_i$, for our reference values of $\sigma_{12}$.}
    \label{fig:nm_ratios_der}
\end{figure}

These results indicate that an evolution mapping relation akin to 
equation~(\ref{eq:pk_evmap}) for the matter power spectrum
is also applicable to $n(M)$, that is
\begin{equation}
n(M|z,\bm{\Theta}_{\rm s},\bm{\Theta}_{\rm e}) \simeq n(M|\bm{\Theta}_{\rm s},\sigma_{12}\left(z,\bm{\Theta}_{\rm s},\bm{\Theta}_{\rm e}\right)).
\label{eq:evmap_nm}
\end{equation}
In the same way as the non-linear matter power spectrum, 
the deviations from this relation can be described in terms of the different structure 
formation histories of these models as 
\begin{equation}
\begin{split}
n\left(M|g, g'\right) = &\,\, n\left(M|g_0\,,\,g'_0\right) + 
\frac{\partial n }{\partial g}\left(M|g_0\,,g'_0\right)\left(g - g_0\right)\\ 
&+ 
\frac{\partial n }{\partial g'}\left(M|g_0\,,g'_0\right)\left(g' - g'_0\right),
\label{eq:nm_taylor}
\end{split}
\end{equation}
where we omitted the dependency on $\bm{\Theta}_{\rm s}$ and $\sigma_{12}$. 
Fig.~\ref{fig:nm_ratios_der} shows the ratio between the mass functions of the 
different Aletheia cosmologies
and that inferred using equation~(\ref{eq:nm_taylor}) with the derivatives with respect to 
$g$ and $g'$ estimated from the results of models 0, 1, and 7. 
This relation gives a good description of the mass functions of all models, with 
the deviations being dominated by the variance of the measurements. 
As for the case of $P(k)$, 
the dependence of the residuals from the exact evolution mapping relation for $n(M)$ of 
equation~(\ref{eq:evmap_nm}) on the structure formation histories of these cosmologies 
can be encapsulated into the differences in the values of $g(\sigma_{12})$ and 
$g'(\sigma_{12})$ at the particular value of $\sigma_{12}$ being considered.

\begin{figure}
	\includegraphics[width=\columnwidth]{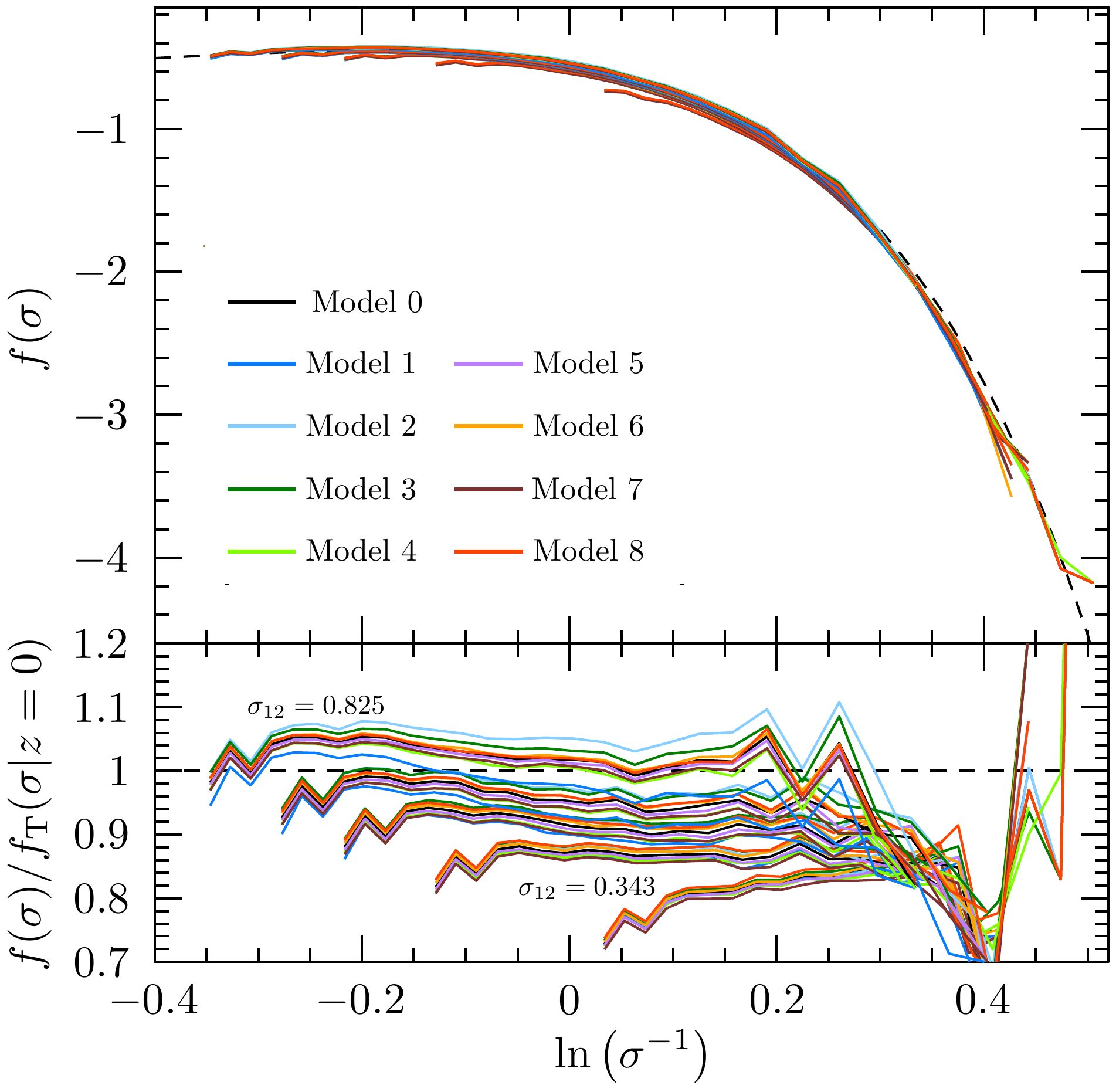}
    \caption{Upper panel: function $f(\sigma)$ defined in equation~(\ref{eq:fsigma}) recovered from the Aletheia simulations
for our reference values of $\sigma_{12}$ (solid lines) compared against the prediction from \citet{Tinker2008} at $z=0$ 
(dashed line). Lower panel: ratio between the simulation results and $f_{\rm T}(\sigma|z=0)$. The deviations from universality 
are more correctly described in terms of $\sigma_{12}$ than the redshift $z$.}
    \label{fig:fsigma}
\end{figure}

Several studies have shown that $f(\sigma)$ cannot be described by a universal function at high accuracy as 
its amplitude and shape are cosmology and redshift dependent
\citep{Tinker2008,Courtin2011, Despali2016, Diemer2020, Ondaro2021}. Fig.~\ref{fig:fsigma} illustrates this result. 
The upper panel shows the function $f(\sigma)$ inferred from the Aletheia simulations for our reference values 
of $\sigma_{12}$ (solid lines), compared against the $z=0$ prediction from \citet{Tinker2008}, 
$f_{\rm T}(\sigma|z=0)$ (dashed lines). 
The amplitude of the halo multiplicity functions inferred from the simulations decreases with increasing redshift. 
These deviations can be seen more clearly in the lower panel, which shows the ratio of the simulation results 
with the theory prediction at $z=0$. 
This plot also shows that, although they are characterized by different redshifts, the functions $f(\sigma)$ 
corresponding to a given value of $\sigma_{12}$ become increasingly similar as this parameter decreases. 
This suggests that, for this mass definition, the key parameter to describe the deviations from universality of $f(\sigma)$ is 
$\sigma_{12}$ and not the redshift explicitly. This can be used to provide improved recipes of the evolution 
of the mass function that 
are accurate at high redshifts, specially for cosmologies that deviate from the standard $\Lambda$CDM 
parameters.

\section{Practical applications of evolution mapping}
\label{sec:applications}

In this section we focus on the practical applications of the results presented in 
Sec.~\ref{sec:evmap} on the analysis and interpretation of numerical simulations 
and cosmological observations.

At the level of the linear evolution of density fluctuations, the deceptively simple relation 
of equation~(\ref{eq:pk_evmap}) has several implications as it drastically reduces the number 
of parameters required to describe $P_{\rm L}(k|z)$. 
For example, when using the Markov Chain Monte Carlo (MCMC) technique to derive
cosmological parameter constraints from LSS data sets, the values of all evolution 
parameters such as $w_0$ or $w_a$ are treated as slow quantities that require a new 
call to a Boltzmann solver such as {\sc camb} \citep{Lewis1999} or {\sc class} \citep{Lesgourgues2011} 
to compute $P_{\rm L}(k|z)$ every time their values are changed. 
The computational cost of these multiple evaluations is a bottleneck of the analysis. 
However, once the power spectrum of a given model has been computed, 
a large number of alternative models defined by the same shape parameters and varying 
evolution parameters can be sampled simply by re-scaling the obtained $P_{\rm L}(k|z)$, 
expressed in Mpc units, by the ratios of their corresponding  values of $\sigma_{12}^2(z)$. 
In essence, the only quantities that should be treated as slow parameters are the physical 
density parameters $\omega_{\rm b}$, $\omega_{\rm c}$, and $\omega_{\nu}$ that 
define the shape of the transfer function. This approach can lead to a significant speed-up of 
current cosmological analyses, allowing us to explore larger parameter spaces in 
considerably less time.  

Recent studies have proposed to reduce the computational cost of parameter inference 
analyses by means of emulators of $P_{\rm L}(k|z)$ calibrated on 
 the outputs of Boltzmann solvers \citep{Arico2021, Mancini2021}. 
These emulators are based on large training sets that sample different cosmological 
parameters including purely evolution parameters, 
such as $w_0$ or $w_a$, and parameters that represent a mixture of evolution and shape 
quantities such as $\sigma_{8/h}$ or the fractional density parameters $\Omega_{\rm c}$
and $\Omega_{\rm b}$.
The calibration of these emulators makes use of $P_{\rm L}(k|z)$ expressed in 
$h^{-1}{\rm Mpc}$ units, which is explicitly evaluated at multiple redshifts. 
A more efficient emulator design can be achieved by exploiting the fact that, 
when expressed in Mpc units, evolution parameters only affect the linear power spectrum 
through the value $\sigma_{12}(z)$.

If the dependence of the transfer function, $T(k)$, on the relevant physical density parameters
is emulated, it is possible to obtain a prediction for the full power 
spectrum in ${\rm Mpc}$ units for any choice of evolution parameters and redshift. 
The full shape of $P_{\rm L}(k|z)$ can be obtained from the output of the emulator of $T(k)$
and $n_{\rm s}$ (and any additional shape parameter controlling the scale dependence of the 
primordial power spectrum). 
The correct amplitude of $P_{\rm L}(k|z)$ can be obtained by computing the value of 
$\sigma_{12}(z)$ in terms of the desired normalization at  $z=0$ and the growth 
factor $D(z)$ for the cosmology being considered. 
Such recipe would be valid for a wide range of evolution cosmological parameters, 
including non-standard models such as EDE or other possible dynamic dark
dark energy cosmologies, as long as the corresponding linear growth factor is scale independent. 
The smaller number of parameters required to emulate $T(k)$ should increase the overall 
accuracy of the predictions, while avoiding the need to sample the redshift evolution 
of $P_{\rm L}(k|z)$ explicitly should greatly simplify the calibration procedure. 

The evolution mapping relation of equation~(\ref{eq:pk_evmap}) can also be used to 
construct more general emulators of the non-linear matter power spectrum. 
Emulators based on the outputs of N-body simulations have now become a common tool 
to describe the non-linear evolution of the matter power spectrum 
\citep{coyote2010, Heitman2016, AemulusI, Garrison2018, Euclidemulator2018, Knabenhans2020}. 
Due to the high computational cost of the required simulations, the number of 
nodes in the emulator design must be kept to a minimum. This makes it difficult to explore a 
high-dimensional cosmological parameter space while maintaining accurate predictions. 
As a result, current emulators leave out parameters such as the curvature of the Universe, 
or dynamic dark energy models beyond the standard parametrization of equation~(\ref{eq:wde}). 
A large number of redshift outputs are also required to probe the evolution of these
cosmologies, which complicates the calibration of the emulators. 
These problems could be alleviated by using an alternative emulator design that 
exploits the mapping between models characterized by the same value of $\sigma_{12}(z)$.
This could be achieved by choosing a reference set of evolution parameters, 
$\bm{\Theta}_{{\rm e},0}$, which is kept fixed in all simulations, and sampling only over the
parameter space 
\begin{equation}
\bm{\Phi} = \left(\bm{\Theta}_{\rm s}, \sigma_{12}\right).
\end{equation}
Each node $\bm{\Phi}_i = \left(\bm{\Theta}_{{\rm s},i}, \sigma_{12,i}\right)$ would
correspond to a simulation with the cosmological parameters 
$\left(\bm{\Theta}_{{\rm s},i},\bm{\Theta}_{{\rm e},0} \right)$ and a single redshift 
output such that 
\begin{equation}
\sigma_{12}(z,\bm{\Theta}_{{\rm s},i},\bm{\Theta}_{{\rm e},0})=\sigma_{12,i}.
\end{equation}
The power spectrum of a cosmology characterized by any set of evolution parameters 
$\bm{\Theta}_{{\rm e}}$ could then be obtained by evaluating the emulator at its
corresponding value of $\sigma_{12}(z)$. This prediction, which would correspond to 
the reference set $\bm{\Theta}_{{\rm e},0}$, could then be transformed to the desired 
evolution parameters using a recipe similar to equation~(\ref{eq:pk_taylor})
to account for the different formation histories of the two models.  
With the exception of this second step, a similar approach was used by 
\citet{Pedersen2021} to emulate the Lyman-$\alpha$ forest one-dimensional power spectrum 
in terms of $\Delta^2_{\rm p}$ 
\citep[see also ][]{McDonald2005}.

The smaller number of parameters involved in the emulation procedure, added to the 
fact that no explicit sampling of $z$ would be required, would result in more accurate 
predictions. Furthermore, such an emulator would be automatically valid for all possible 
choices of evolution parameters. We are currently producing an emulator that will serve 
as a proof of concept of this design. This emulator, which we call {\sc Cassandra}, 
will be described in more detail in a forthcoming publication (Gonzalez-Jara et al., in prep.).

The analysis of the halo mass function of Sec.~\ref{sec:nm} suggests that the same 
general strategy can be applied to model other statistics of the non-linear density field. 
Using a reference set of evolution parameters, it is possible to study the 
evolution of the desired statistic as a function of $\sigma_{12}$. Those models could then 
be extended to describe general cosmologies by analysing the impact of the different 
structure formation histories, characterized by $g(\sigma_{12})$, on the results.

\section{Conclusions}
\label{conclusions}

We analysed the impact of different cosmological parameters on the 
matter power spectrum and on structure formation in general. 
We classified all parameters into two sets, shape and evolution parameters.
The former are parameters that affect the shape of the linear-theory dimensionless 
power spectrum.  The latter include parameters that only determine the time 
evolution of the amplitude of $\Delta^2_{\rm L}(k|z)$. 
The effect of all evolution parameters is degenerate:
once the shape parameters are fixed, the linear dimensionless power spectra of models with 
different evolution parameters that result in the same clustering amplitude,  
even if this occurs at different redshifts, are identical. 

As the linear-theory power spectrum is the key quantity to determine the  
properties of the non-linear density field,  the  evolution mapping relation of 
equation~(\ref{eq:pk_evmap}) can be used to reduce significantly 
the number of parameters required to model the impact of non-linearities on $P(k|z)$.
We tested this relation using the Aletheia simulations, a set of paired-fixed 
N-body simulations of nine cosmologies defined by the same shape parameters
and a wide array of evolution parameters (summarized in Table~\ref{tab:all_cases}). 
When measured at the redshifts for which their respective values of $\sigma_{12}(z)$
are identical, the power spectra of all Aletheia cosmologies are remarkably similar and 
only show differences of a few percent in the deeply non-linear regime. The residuals with 
respect to the exact evolution mapping relation are the result of the different 
structure formation histories experienced by each model to 
reach the same value of $\sigma_{12}(z)$. For the range of scales 
included in this analysis, these differences can be accurately described in terms of the 
values of $\Delta g(\sigma_{12})$ and $\Delta g'(\sigma_{12})$ between 
these cosmologies.

The degeneracy between all evolution parameters that lead to the same 
overall clustering amplitude does not only apply to the matter power
spectrum. The full density field inferred from the Aletheia simulations
corresponding to the same values of $\sigma_{12}(z)$ are remarkably 
similar. This suggests that a relation analogous  to 
equation~(\ref{eq:pk_evmap}) could be valid in general for all 
statistical descriptions of the density field. As an example, we 
showed that the halo mass functions of the Aletheia cosmologies 
are in good agreement, with residuals of a few per cent that can 
be described in terms of their differences in $g(\sigma_{12})$ and 
$g'(\sigma_{12})$ using the same general procedure as for $P(k|z)$.

We discussed a few practical applications of the evolution mapping relation.
At the linear level, this relation can help to significantly speed up the Bayesian analysis 
of LSS data sets that require an evaluation of $P_{\rm L}(k|z)$ for multiple cosmologies
by treating all evolution parameters as fast quantities that do not require a 
new call to a Boltzmann solver. It can also be used to construct general predictions 
of $P_{\rm L}(k|z)$ based on an emulator of the dependence of $T(k)$ on the shape parameters. 
We also proposed a new design for an emulator of the non-linear power spectrum
in which only the shape parameters are explicitly sampled and all evolution parameters
are implicitly included through different values of $\sigma_{12}$. 
The  predictions of this emulator can be adapted to an arbitrary choice of 
evolution parameters using a recipe similar to equation~(\ref{eq:pk_taylor})
to account for the different structure formation histories between the 
reference cosmology and the desired model.  
Preliminary tests show that an emulator based on this design can provide 
high accuracy prediction of the non-linear power spectrum for 
general cosmologies (Gonzalez-Jara et al., in prep.). The same strategy
can be applied to other statistics of the density field and can be used as 
a general approach to describe matter clustering in the non-linear regime.

Although the values of the basic shape parameters are well constrained by present-day CMB 
observations, our current constraints on evolution parameters still allow for significant deviations
from the standard $\Lambda$CDM model. It is important to obtain general observational 
constraints on the evolution of $\sigma_{12}(z)$ as it encapsulates the information on the 
values of all evolution parameters. 
Measurements of $\sigma_{12}(z)$ within a particular parameter space also impose strong 
constraints on the shape of the matter power spectrum in the non-linear regime that can 
be used to extend current galaxy clustering analyses into smaller scales. 

The evolution mapping relations discussed here only become evident when the power spectra 
and halo mass functions are expressed in Mpc and ${\rm M}_\odot$ units and 
cannot be described in terms of the commonly used $\sigma_{8/h}$. 
Our findings complement the arguments of \citet{Sanchez2020} against the use of the 
traditional factors of $h$ that are commonly introduced in the units of theoretical predictions for 
cosmological observables. There is much to be gained by abandoning this practice and expressing 
models of $P(k|z)$, and the results of cosmological numerical simulations in general, in Mpc 
units.

\section*{Acknowledgements}

We would like to thank Daniel Farrow, Jiamin Hou, Martha Lippich,  Andrea Pezzotta, 
Agne Semenaite, Mart\'in Crocce, Alex Eggemeier, Rom\'an Scoccimarro, Benjam\'in 
Camacho, and Cristi\'an S\'anchez for their help and useful discussions. 
This work was performed during the COVID-19 pandemic with prolonged 
lockdowns. We would like to thank all health care providers and other 
essential workers for their invaluable efforts during this crisis. 
This research was supported by the Excellence Cluster ORIGINS, 
which is funded by the Deutsche 
Forschungsgemeinschaft (DFG, German Research Foundation) under 
Germany's Excellence Strategy - EXC-2094 - 390783311.
JGJ acknowledges support from CONICYT/ANID-PFCHA/Doctorado Nacional/2021-21210846 and the CONICYT Basal project AFB-170002.
The Aletheia simulations were carried out and post-processed on the HPC system 
Raven of the 
Max Planck Computing and Data Facility (MPCDF) in Garching, Germany. 

\section*{Data Availability}

The data underlying this article will be shared on reasonable request to the corresponding authors.



\bibliographystyle{mnras}
\bibliography{evolution_mapping}




\appendix
\section{Comparison with available prediction schemes of the non-linear
power spectrum}
\label{sec:appendix}

In this section, we test the ability 
of current recipes of the non-linear $P(k|z)$ to reproduce the relation between models with identical 
values of $\sigma_{12}(z)$ described in Section~\ref{sec:evmap_nonlin}. 
We focus on two such recipes, namely, (i) the {\sc HMcode} prescription \citep{Mead2015,Mead2016,Mead2021} as implemented 
in {\sc camb}, and (ii) on the {\sc EuclidEmulator 2} \citep{Knabenhans2020}. 
As these recipes have not been designed to account for the evolution-mapping 
relation of equation~(\ref{eq:pk_evmap}), it is interesting to test how well they 
can reproduce the ratios of the power spectra of the different Aletheia simulations.

\begin{figure}
	\includegraphics[width=\columnwidth]{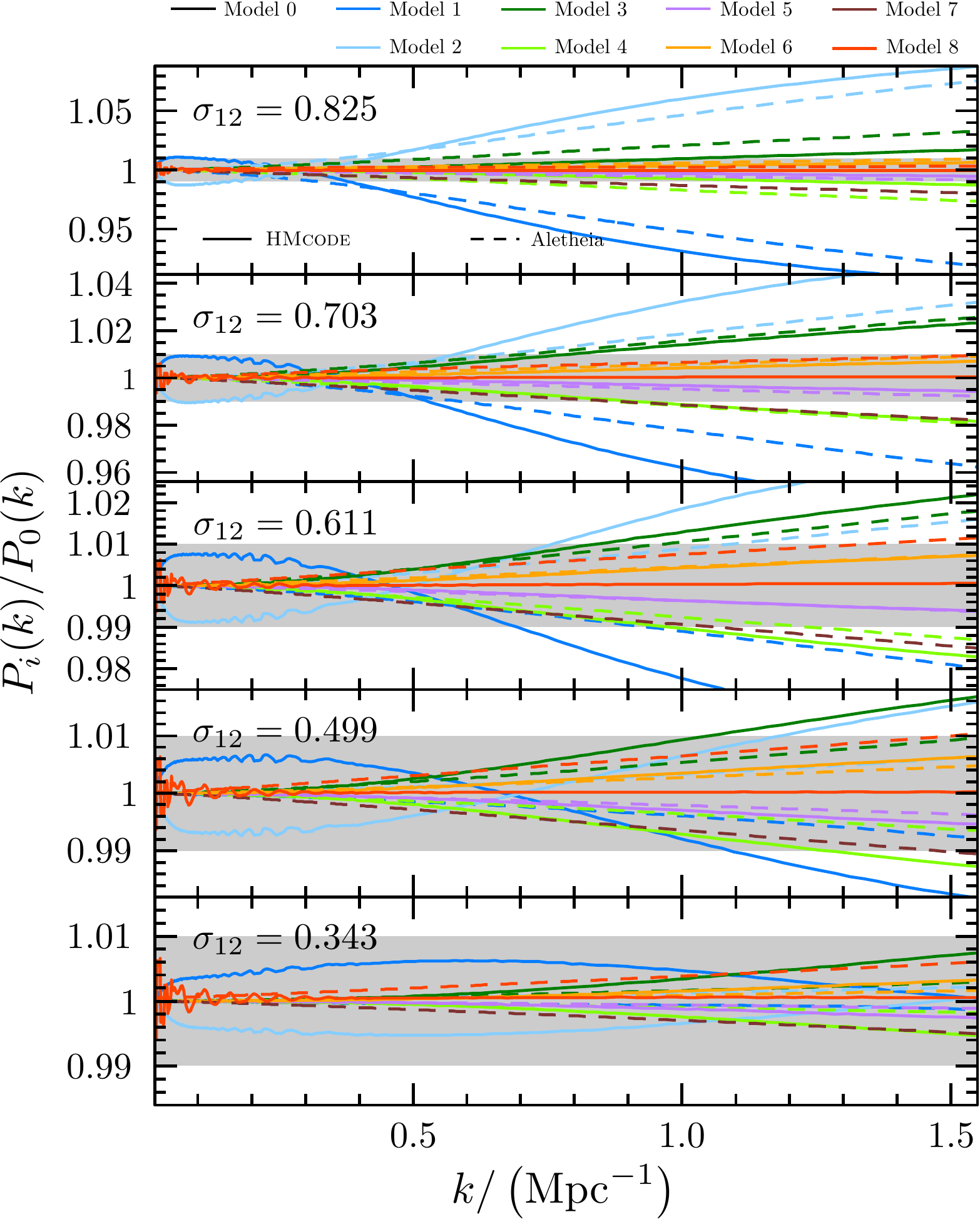}
    \caption{Ratios of the power spectra of the test cosmologies defined in 
    Table~\ref{tab:all_cases} with that of model 0 for our reference values of $\sigma_{12}$ 
    as predicted by {\sc HMcode} (solid lines). For comparison,
    the dashed lines show the results obtained from the Aletheia simulations (matching those shown in 
    Fig.~\ref{fig:pk_ratios}).
    }
    \label{fig:ratios_hmcode}
\end{figure}

\begin{figure}
	\includegraphics[width=\columnwidth]{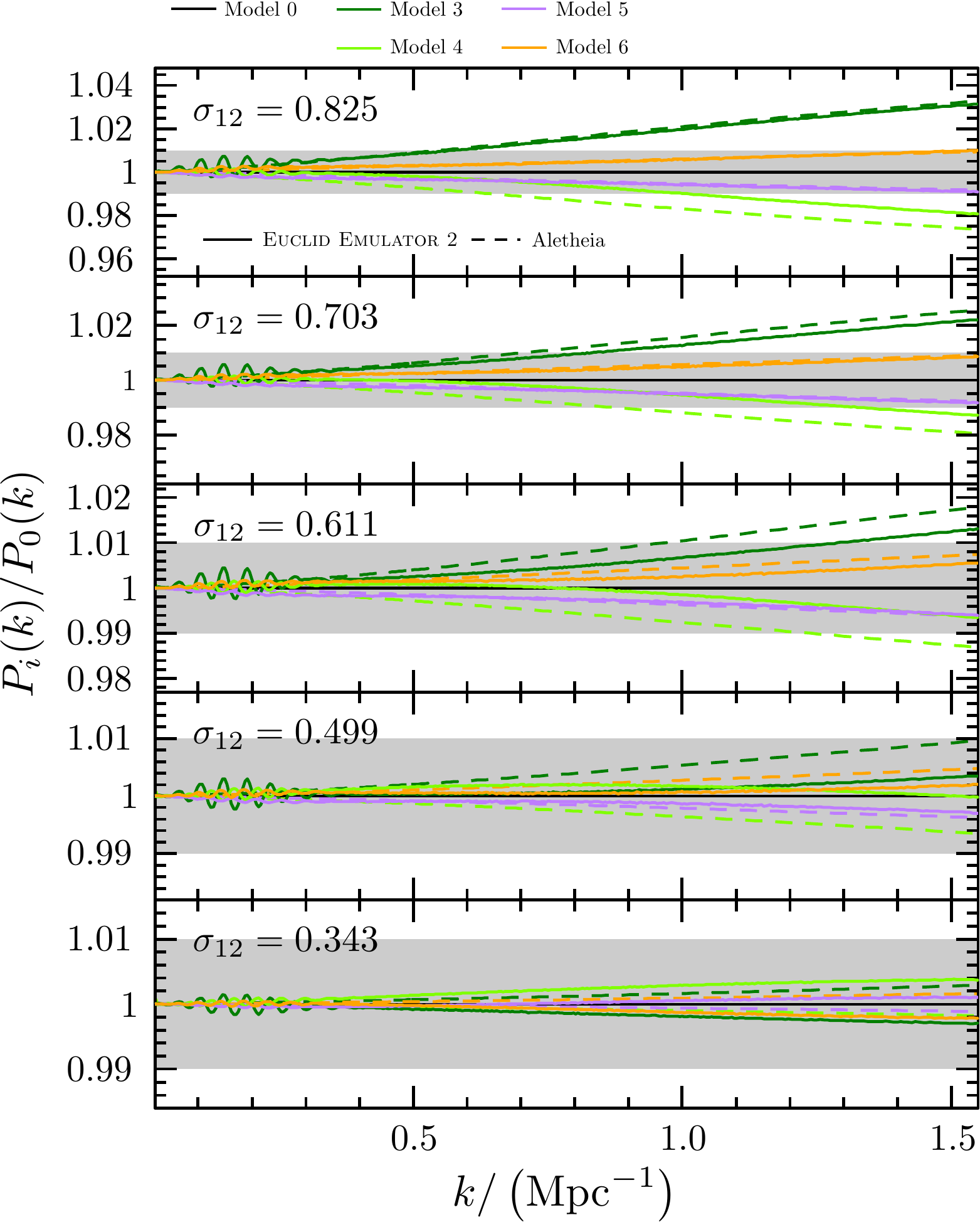}
    \caption{ 
    Same as Fig.~\ref{fig:ratios_hmcode} but based on the predictions of the {\sc EuclidEmulator 2} (solid lines). Only 
    the models that fall within the allowed parameter ranges of the emulator are included. The dashed lines show 
    the results of the corresponding Aletheia simulations (also shown in Fig.~\ref{fig:pk_ratios}).
     }
    \label{fig:ratios_ee2}
\end{figure}

The recipe of {\sc HMcode} is based on the halo model of structure formation 
\citep{Ma2000,Seljak2000, COORAY2002} and produces power spectra predictions
that are accurate at the 5 per cent level for a wide range of non-standard cosmologies. 
The solid lines in Fig.~\ref{fig:ratios_hmcode} show the predictions obtained 
with  {\sc HMcode} of the ratios of the power spectra of all Aletheia cosmologies 
with that of model 0 for our reference values of $\sigma_{12}$. For comparison, 
the results inferred from the 
Aletheia simulations are shown by the dashed lines.
In general, {\sc HMcode} gives an adequate description of the results of the simulations, 
with differences at the level of a few per cent within the range of scales considered here. 

Rather than emulating $P(k|z)$ directly, the {\sc EuclidEmulator 2} predicts the boost factor, 
$B(k)$, defined as the ratio
\begin{equation}
B(k|z) = \frac{P(k|z)}{P_{\rm L}(k|z)}.
\end{equation}
As for each reference value of $\sigma_{12}$ the linear-theory power spectra of our 
test cosmologies are identical, the ratio of the boost factors predicted by the emulator 
corresponds to the ratio of $P(k|z)$. 
Of all the cosmologies defined in Table~\ref{tab:all_cases}, only models 0 and 3 -- 6
fall within the parameter ranges of the {\sc EuclidEmulator 2}. 
The solid lines in Fig.~\ref{fig:ratios_ee2} show the ratios of the power 
spectra of these cosmologies with that of model 0 as predicted by the emulator. 
The dashed lines show the results obtained from the Aletheia simulations.
The emulator gives an accurate description of the simulation results, with sub-percent 
level deviations in all cases. This agreement suggests that a relation similar to 
equation~(\ref{eq:pk_taylor}) could help to extend the applicability of the emulator outside 
its current parameter range.

\bsp	
\label{lastpage}
\end{document}